\documentclass[letterpaper]{article} 
\usepackage{aaai25}  
\usepackage{times}  
\usepackage{helvet}  
\usepackage{courier}  
\usepackage[hyphens]{url}  
\usepackage{graphicx} 
\usepackage{natbib}  
\usepackage{caption} 
\usepackage{style}
\usepackage{newfloat}
\usepackage{listings}
\DeclareCaptionStyle{ruled}{labelfont=normalfont,labelsep=colon,strut=off} 
\floatstyle{ruled}
\newfloat{listing}{tb}{lst}{}
\floatname{listing}{Listing}
\title{Heterogeneous Multi-Robot Graph Coverage with Proximity and Movement Constraints\thanks{
The research of Yonatan Aumann is supported in part by ISF grant 3007/24. The research of Sarit Kraus is supported in part by ISF grant  2544/24.}}
\author{
    Dolev Mutzari\quad
    Yonatan Aumann\quad
    Sarit Kraus
}
\affiliations{
    Department of Computer Science, Bar Ilan University, Ramat Gan, Israel\\

    dolevmu@gmail.com,
    aumann@cs.biu.ac.il,
    sarit@cs.biu.ac.il
}

\begin{document}

\maketitle

\begin{abstract}
Multi-Robot Coverage problems have been extensively studied in robotics, planning and multi-agent systems. In this work, we consider the coverage problem when there are constraints on the proximity (e.g., maximum distance between the agents, or a blue agent must be adjacent to a red agent) and the movement (e.g., terrain traversability and material load capacity) of the robots.
Such constraints naturally arise in many real-world applications, e.g. in search-and-rescue and maintenance operations. Given such a setting, the goal is to compute a covering tour of the graph with a minimum number of steps, and that adheres to the proximity and movement constraints.   
For this problem, our contributions are four: (\romannumeral 1) a formal formulation of the problem, (\romannumeral 2) an exact algorithm that is FPT in parameters $\|\formations\|$, $\maxdegree$ and $\omega$ - the set of robot formations that encode the proximity constraints, the maximum nodes degree, and the tree-width of the graph, respectively, (\romannumeral 3) for the case that the graph is a tree: a PTAS approximation scheme, that given an $\varepsilon$ produces a tour that is within a $1+\varepsilon\cdot \error(\|\formations\|,\maxdegree))$ of the optimal one, and the computation runs in time $\poly(n) \cdot h(\frac{1}{\varepsilon},\|\formations\|)$.
(\romannumeral 4) for the case that the graph is a tree, with $k=3$ robots, and the constraint is that all agents are connected: a PTAS scheme with multiplicative approximation error of $1+\calO(\varepsilon)$, independent of $\maxdegree$.

\end{abstract}

%

\section{Introduction}
Multi-robot graph-coverage (MRGC) models a multitude of real-world robotic scenarios, including surveillance~\cite{scherer2020multi,zhang2020autonomous,vallejo2020multi,lee2021upper,gans2021cooperative,lee2023efficient}, cleaning applications~\cite{nemoto2020heterogeneous,miao2020multi}, environmental monitoring~\cite{wang2023development}, search and rescue operations~\cite{queralta2020collaborative,rodriguez2020wilderness,yang2020needs,drew2021multi}, warehouse automation~\cite{salzman2020research,bolu2021adaptive,zaccaria2021multi,zhang2024multi}, and agricultural field management~\cite{govindaraju2023optimized,choton2023optimal,mukhamediev2023coverage}, where robots are tasked with covering a defined graph-like structure efficiently and effectively~\cite{galceran2013survey}. In the \emph{cleaning} scenario, for example, a set of \emph{cleaning robots} is tasked with cleaning some environment, such as an office building, visiting (and cleaning) every room in the building. If every robot is independent, this is a simple instance of a multi-robot graph coverage.  In many cases, however, robots act as a \emph{team}~\cite{nemoto2020heterogeneous}, possibly with different specializations.  For example, some robots may be capable of the actual \emph{cleaning}, while other robots may be \emph{carrier robots} - engineered to carry large loads of water or rubbish. In this case, each site/room must be visited by a cleaning robot, but the carrier robots must always be in close proximity.  Similarly, in search and rescue operations, some robots may have searching capabilities, while other \emph{rescue robots} (e.g., diggers, medical, etc.) must tag along and be available in close proximity. Even if all agents are of the same type, communication requirements may constrain them to stay within small proximity of each other.  Additionally, graph edges (e.g. doors in a building) may display different physical properties (e.g., width), constraining the passage of the different robots along different edges (e.g., carrier robots cannot traverse some doors, while cleaning can). In all, graph coverage in such team settings may impose additional constraints on the maximum distance between agents, the permissible formations, and transitions.

In this paper, we consider this \emph{team} multi-robot graph coverage problem. Specifically, we address minimizing the number of steps to cover an input graph with a team of heterogeneous robots (/agents), given constraints on (i) the permissible formations of the agents, (ii) the permissible transitions between agent formations, and (iii) the passage of different agents along graph edges. To the best of our knowledge, no previous work has addressed this setting. Notably, even if the graph is a tree and there are no constraints on agent formations, minimizing the number of steps for coverage is NP-hard~\cite{fraigniaud2006collective}.

\paragraph{Our Contributions.}
Our contributions are four.  We provide (i) a formal formulation of the problem, (ii) an exact algorithm that is FPT in parameters $\maxdegree$, $\tw$, and $||{\formations}||$, respectively: the maximum node degree in the graph, the tree-width of the graph, and the size of the representation of the constraints, (iii) for the case that the graph is a tree: a PTAS approximation scheme, that given an $\varepsilon$ produces a tour that is within a $1+\varepsilon\cdot \error(||{\formations}||,\maxdegree))$ of the optimal one (where $\error(\cdot)$ is independent of the graph size), and the computation runs in time $\poly(n) \cdot h(\frac{1}{\varepsilon},k)$.
(iv) for the case that the graph is a tree, the only constraint is that all agents are connected, and there are three agents: a PTAS scheme with approximation error of $1+\calO(\varepsilon)$, independent of $\maxdegree$.  

\section{Related Work}
We identify a variety of works that are related to multi-robot coverage problems with constraints. Nevertheless, we did not find previous work on the exact problem of interest, that is, multi-robot coverage of graphs (with bounded treewidth) under proximity or connectivity constraints.

In~\cite{fraigniaud2006collective}, it is shown that the \emph{Multi-Robot Connected Tree Coverage (MRCTC)} is NP-hard. However, the parameterized complexity is not analysed. A follow-up work by~\cite{cabrera2012flooding} also considered MRCTC and, in addition, restricted the number of robots allowed to traverse an edge and enter a vertex during each step. Nevertheless, they also did not consider proximity constraints. Instead, coordination is achieved by dropping landmarks at explored vertices, enabling decentralized exploration. However, this approach has drawbacks: (i) landmarks incur costs; (ii) in rescue scenarios, they may be unavailable in time or quantity; (iii) placing them can take additional time. Later, \cite{sinay2017maintaining} proposed an algorithm for MRCTC and focused on connectivity constraints. Since the problem is NP-hard, they focused on the \emph{speedup factor}, that is, the ratio between the multi-robot and the single-robot traversal time. However, no comparison with the optimal solution is provided, and only trees are considered. In~\cite{charrier2020complexity}, the complexity of multi-agent path finding (MAPF) and multi-robot coverage is analysed for topological graphs $G=(V,E_m,E_c)$ with \emph{movement edges} and \emph{communication edges}, where robots must stay connected to a base station. The main results are negative, suggesting both problems are PSPACE-complete.

A well-studied use case for multi-robot coverage is mapping and model reconstruction (see~\cite{almadhoun2019survey} for a recent survey). In these settings, the environment is unknown in advance. \cite{brass2011multirobot} considers Multi-Robot \emph{Unknown Graph} Coverage, and focuses on exploration.
In~\cite{banfi2016asynchronous} the robots must connect to a base station only when information is collected, allowing robots to disconnect for arbitrarily long periods.
We view this line of work as complementary to ours. First, robots can explore and map an environment, but from that point on, we may assume the graph is given as input, e.g. for patrolling.

Another closely related area to our work is Multi-Robot Coverage Path Planning (mCPP), which involves using multiple robots to scan a continuous planar environment. Studies like \cite{tang2021mstc} and \cite{lu2023tmstc} explore mCPP under physical constraints, similar to our work, but they do not address proximity constraints. \cite{jensen2018online} compares mCPPs with varying communication levels. \cite{mechsy2017novel} considers a tethered robot CPP problem, where the robot has a chain structure with a constrained length.
Additionally, Multi-Agent Path Finding (MAPF)~\cite{erdem2013general} focuses on planning non-colliding paths for multiple robots, and \cite{dutta2019multi} examines informative path planning with continuous connectivity constraints.

Proximity and connectivity are vital when considering robot \emph{swarms}. Both~\cite{panerati2018swarms} and~\cite{siligardi2019robust} study the problem of maintaining swarm connectivity while
performing a coverage of an area of interest.  In (Liu et al. 2023), land is scanned by UAV swarms with limited perception, while (Tran et al. 2023) extends this to repeated coverage with heterogeneous robots for dynamic environments. While swarm robotics is designed to scale with the number of robots, the model is somewhat limited. Indeed, robots cannot fully coordinate; they must follow relatively simple rules based on local observation and local communication, and decisions are made in real-time, individually, and asynchronously. A centralized planner, despite its limitations, can provide more efficient coverage.

\section{Multi-Robot Connected Graph Coverage}
Consider an undirected graph $G=(V,E)$, parameterized by its treewidth $\tw$ and maximal degree $\maxdegree$. An edge $e_{ij}=\{v_i,v_j\}$ exists if a robot can move directly from $v_i$ to $v_j$.

A \emph{configuration} of robots $\xx: V \rightarrow \NN$ specifies how many robots occupy each vertex. In a \emph{connected} configuration, the set of \emph{occupied} vertices $\occupied{\xx}:=\{v \in V: x(v) > 0\}$ forms a connected sub-graph of $G$. A simple interpretation is line-of-sight. A \emph{transition} is a pair of connected configurations $(\xx,\xx')$, where $\xx'$ can be reached from $\xx$ by moving each robot along an edge. A $t$-\emph{traversal} of $T$ is a sequence of $t+1$ connected configurations $\calX=(\xx^0,\ldots,\xx^t)$, that form $t$ sequential transitions, where each vertex $v\in V$ is visited at least once by at least one robot.

The \emph{Multi-Robot Connected Graph Coverage ($\MRCGC$)} Problem is defined as follows: Given a graph $G$, a number of robots $k \in \NN$ initially located at an entry point $s \in V$, find a \emph{traversal} $\calX$ of minimal time $\time{\calX}:=|\calX|=\timeoptimal$ that starts and terminates with all robots at $s$.\footnote{We identify traversal time with traversal length. In practice, long edges can be split into shorter edges by inserting vertices.}

\section{Multi-Robot Formation Graph Coverage}
In this work, we study an extension of $\MRCGC$ that considers a finite set $M$ of robot \emph{types}. In the heterogeneous setting, for each $m\in M$, there are $k_m$ robots, and $k=\sum_{m}{k_m}$ robots overall. A configuration is now given as $\xx: V\times M\rightarrow \NN$, stating for each vertex $v\in V$, and for each robot type $m\in M$, how many robots of type $m$ occupy $v$.

The set of valid configurations $\configs$ is extended as well by considering \emph{formations}. A \emph{formation} of robots is a pair $\alpha=\langle G_\alpha, \xx_\alpha \rangle$ where $G_\alpha=(V_\alpha,E_\alpha)$ is a connected, undirected graph, and $\xx_\alpha: V_\alpha \times M \rightarrow \NN$ is a configuration of $k$ robots on $G_\alpha$. A formation represents a valid way of positioning the robots. We say that \emph{$\xx$ is in $\alpha$-form} if there exists a \emph{graph monomorphism} $\phi: V_\alpha \rightarrow V$ such that $\xx(\phi(v),m)=\xx_\alpha(v,m)$ for each $v\in V_\alpha$ and each $m\in M$. We call $\activated{\xx}:=\phi(V_\alpha)$ the set of \emph{active} vertices in configuration $\xx$. Note that $\occupied{\xx} \subseteq \activated{\xx}$. However, in general unoccupied vertices may be active as well. The set of valid configurations $\configs$ is then dictated by the set of formations $\formations=\{\langle G_\alpha, \xx_\alpha \rangle\}_\alpha$. We denote by $\|\formations\|$ the representation length of the set $\formations$, where graphs are represented with adjacency lists and maps are stored as $V_\alpha \times M$ tables.

Lastly, transitions are restricted by considering \emph{transpositions}. A transposition is a pair $\langle G_{\{\alpha,\alpha'\}}, \{\xx_\alpha, \xx_{\alpha'}\} \rangle$, where $\xx_\alpha,\xx_{\alpha'}$ are configurations in $G_{\{\alpha,\alpha'\}}$ of $\alpha,\alpha'$ form respectively, where $\xx_{\alpha'}$ can be reached from $\xx_\alpha$ by moving each robot along up to one edge in $G_{\{\alpha,\alpha'\}}$. A transposition represents one possible valid way of moving the robots. We say that transition $(\xx,\xx')$ is in $(\alpha,\alpha')$ form if $\xx,\xx'$ are in $\alpha,\alpha'$ form respectively. The set of valid transitions is determined by the set of transpositions $\transpositions$. We refer to Appendix~\ref{sec:examples} for illustrative examples.

The \emph{Multi-Robot Formation Graph Coverage ($\MRFGC$)} Problem is defined as follows: Given a Graph $G$, a set $\formations$ of formations, a set $\transpositions$ of transpositions, a start configuration $\xx_0$ and an end configuration $\xx_f$, find a \emph{traversal} $\calX$ of minimal time that starts at $\xx_0$ and ends at $\xx_f$.

\section{Z-Lemma: Transitions do not Repeat}
In this section, we prove that in an optimal traversal a transition cannot repeat. The proof technique is the cornerstone that enables all follow-up results presented in this work.


\begin{Lemma}
\label{lemma:z-method}
If $\calX$ is optimal, then no transition repeats.
\end{Lemma}

\begin{proof}
Assume in contradiction that some transition repeats. That is, there exist $i<i'$ such that $\xx^{i}=\xx^{i'}$ and $\xx^{i+1}=\xx^{i'+1}$. Then we can construct a shorter traversal, denoted (of course) $\calZ$ in contradiction, as depicted in Figure~\ref{fig:z-lemma}.

\begin{figure}[htb]
\centering
\includegraphics[width=0.8\linewidth]{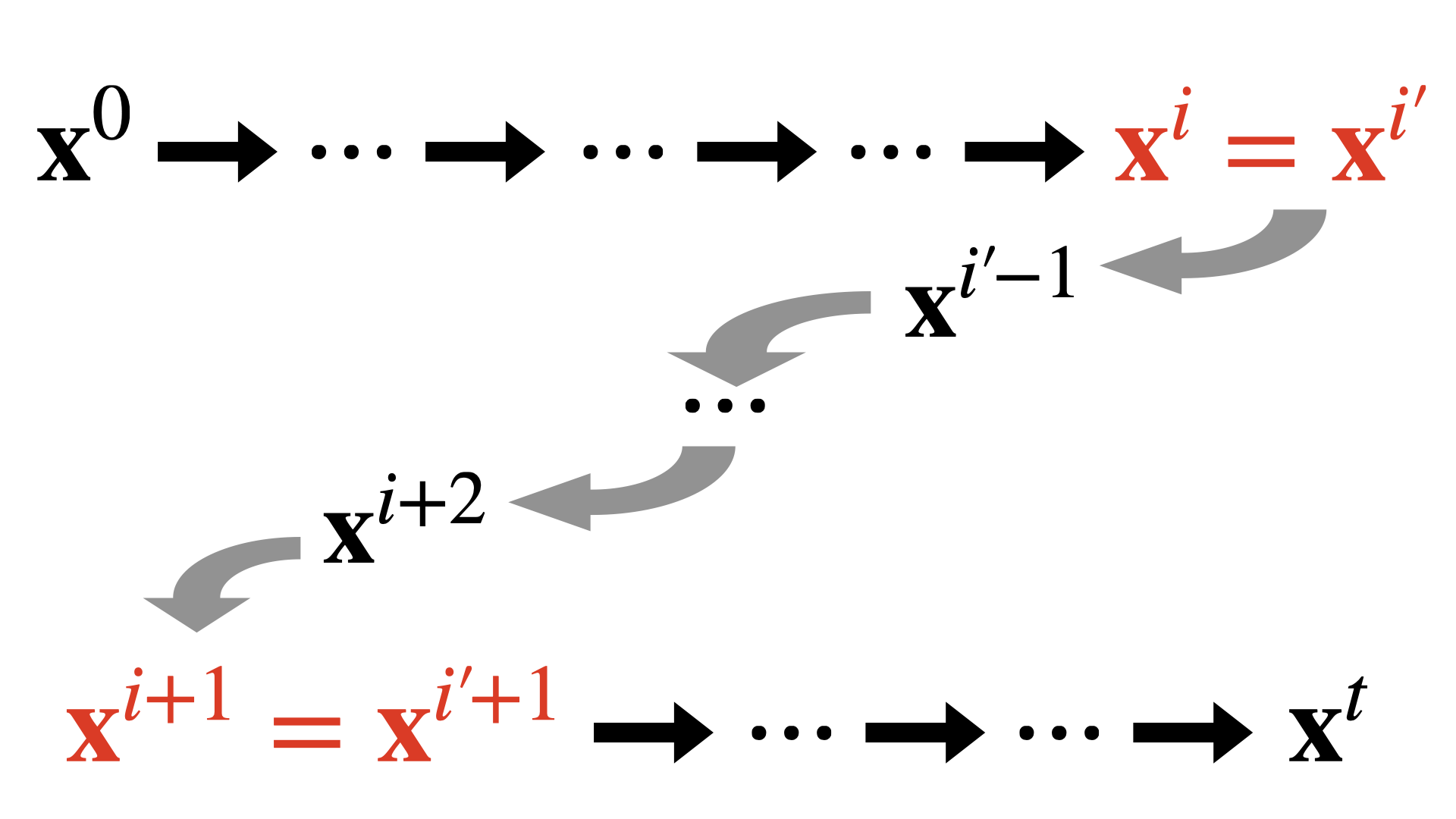}
\caption{Traversal $\calZ$. The repeated configurations are highlighted in red, flipped transitions are colored in gray.} \label{fig:z-lemma}
\end{figure}


Traversal $\calZ$ is formally defined as follows. It follows $\calX$ from initial configuration $\xx^{0}$ up until $\xx^{i}$. By assumption, this is the same configuration as $\xx^{i'}$. From there, traversal $\calZ$ follows $\calX$ \emph{in reverse}, to configuration $\xx^{i'-1}$, and follows this sequence until reaching configuration $\xx^{i+1}$. Here we critically rely on the transpositions being undirected, so that whenever $(\xx,\xx')$ is a valid transition (of some form $(\alpha,\alpha')$), so is $(\xx',\xx)$ (in form $(\alpha',\alpha)$). Finally, as $\xx^{i+1}=\xx^{i'+1}$, traversal $\calZ$ follows $\calX$ from this point onward.

Hence, $\calZ$ is composed of valid transitions. It also covers $G$, as it consists of the same set of configurations as $\calX$. However, traversal $\calZ$ avoids repeating $\xx^i=\xx^{i'}$ and $\xx^{i+1}=\xx^{i'+1}$, and so $\time{\calZ}=\time{\calX}-2$, a contradiction.
\end{proof}

\section{Solving {$\MRFGC$} in $\FPT$-time}
In this section, we develop an algorithm for $\MRFGC$ that is $\FPT$ in treewidth $\tw(G)$, maximal degree of the graph $\maxdegree$, and $\|\formations\|$. We use a bottom-up dynamic programming approach on a \emph{nice tree decomposition~\cite{kloks1994treewidth}} (see Appendix~\ref{sec:nice-tree-decomp}) $\calT$ of $G$. It recursively computes a \emph{table of signatures} at each node of $\calT$, starting from the leaves. In our case, a signature at some node $j$ is a sequence of robot configurations that visit the corresponding bag $B_j\in\calB$, separated by special $\uparrow$ and $\downarrow$ formal symbols that encode that the robots have ``left'' the bag. Due to Lemma~\ref{lemma:z-method}, the sequence length is independent of graph size $|V|$ (Lemma~\ref{lemma:enumerate-sigs}). We show we can recursively update the $\emph{cost}$ of each signature bottom-up, and also backtrack a traversal from a signature at the root, by maintaining back-pointers to signatures of children in $\calT$.

Let $(\calB,\calT)$ be a nice tree decomposition of $G$. Here $\calB=\{B_j\}_{j\in J}$ is the set of bags ($B_j \subseteq V, |B_j|\le \tw+1$), and $\calT$ is the tree structure over $\calB$. For $B\in \calB$, denote by $\configs_{B}$ the set of configurations whose set of active vertices intersects $B$.
Let $V_\downarrow (j) = \cup_{j'\in \calT_j}{B_{j'}} \setminus B_j$ be the vertices that appear solely in the bags under $B_j$. Similarly, let $V_\uparrow (j) = V \setminus V_\downarrow (j) \setminus B_j$ be the vertices that appear solely not in or under $j$.

\begin{Definition}
\label{def:projection}
Let $\calX=(\xx^0,\ldots,\xx^t)$ be a traversal and fix a bag $j\in J$. The \emph{projection of $\calX$ on $j$}, denoted $\project{\calX}{j}$, is the sequence $\calY=(\yy^i)_{0\le i\le t} \in \configs_{B_j} \cup \{\uparrow,\downarrow\}$, where for each $i$:

\begin{align*}
\yy^i = \begin{cases}
    \xx^i      & \activated{\xx^i} \cap B_j \neq \emptyset \\
    \uparrow   & \activated{\xx^i} \subseteq V_\uparrow (j) \\
    \downarrow & \activated{\xx^i} \subseteq V_\downarrow (j) \\
\end{cases}
\end{align*}

The \emph{condensed form} of $\project{\calX}{j}$, denoted $\cproject{\calX}{j}$, is obtained from $\project{\calX}{j}$ by replacing any consecutive repetition of the same element with a single occurrence of that element.
\end{Definition}

\begin{Lemma}
\label{lemma:identify-patterns}
Let $\calX$ be an optimal traversal and $j\in J$. Then $\cproject{\calX}{j}=(\cproject{\xx}{j}^0,\cproject{\xx}{j}^1,\ldots)$ admits the following:
\begin{enumerate}
    \item \label{cond:seq-domain} If $\cproject{\xx}{j}^i \not\in \{\uparrow,\downarrow\}$, then $\cproject{\xx}{j}^i \in \configs_{B_j}$.
    \item \label{cond:seq-trans-repeats} If $\cproject{\xx}{j}^i=\cproject{\xx}{j}^{i'}\in\configs_{B_j}$, and $\cproject{\xx}{j}^{i+1}=\cproject{\xx}{j}^{i'+1}$, then $\cproject{\xx}{j}^{i+1}\not\in\configs_{B_j}$ (\emph{no transition repeats}).
    \item \label{cond:seq-trans-valid} If $\cproject{\xx}{j}^i,\cproject{\xx}{j}^{i+1} \in \configs_{B_j}$, then it is a transition.
\end{enumerate}
\end{Lemma}

Denote by $\poss{j}$ the set of condensed sequences over $\configs_{B_j}\times \{\uparrow, \downarrow\}$, for which 1-3 of Lemma~\ref{lemma:identify-patterns} hold.

\begin{Lemma}
\label{lemma:enumerate-sigs}
There exists an algorithm $\textsf{enumerate\_patterns}$ and function $h(\|\formations\|,\maxdegree,\tw)$ such that given: a graph $G$, a tree decomposition $(\calB,\calT)$ of $G$ with treewidth $\tw$, and a bag $j\in J$, enumerates $\poss{j}$ in time $h(\|\formations\|,\maxdegree,\tw)$.
\end{Lemma}

\begin{proof}
Let $\calY$ be a condensed sequence satisfying 1-3. We first bound the length of the sequence $\calY$. Let $v\in B_j$, and let $I_v$ be the set of indices $i$ where $\yy^i \in \configs_v$, that is, $v\in \activated{\yy^i}$.
The number of such possible configurations in some $\alpha$ form is bounded by some $f_0(\|\langle G_\alpha \rangle \|,\maxdegree)=\maxdegree^{|V_\alpha|-1}$, since formations are represented in explicit form, and the graph monomorphism maps neighbors to neighbors. Therefore, there are up to $f(\|\formations\|,\maxdegree)=\sum_\alpha {f_0(\|\langle G_\alpha \rangle \|,\maxdegree)}\le |\formations| \cdot \maxdegree^{\max_\alpha {|V_\alpha|}}$ possible configurations where $v$ is activated overall. Then, the number of possible transitions from a given configuration with $v$ activated can be similarly bounded by $g_0(\|\formations\|,\maxdegree)=\binom{|\formations|}{2} \cdot \maxdegree^{\max_{\alpha}{|V_{\alpha}|}}$. Fix $g:=2g_0$ to account for transitions where $v$ is activated in the second configuration. Then by the pigeon-hole principle, if $|I_v|>f(\|\formations\|,\maxdegree) \cdot g(\|\formations\|,\maxdegree)$, there is a transition that repeats, violating Condition~\ref{cond:seq-trans-repeats}. Therefore, $|\bigcup_{v\in B}{I_v}| \le f(\|\formations\|,\maxdegree) \cdot g(\|\formations\|,\maxdegree)\cdot (\tw+1)$.

Therefore, the length of a condensed sequence is bounded by $f(\|\formations\|,\maxdegree)\cdot g(\|\formations\|,\maxdegree)\cdot (\tw+1)$, and it is over an alphabet of size $f(\|\formations\|,\maxdegree)\cdot (\tw+1)+2$. An exhaustive search may enumerate over all such sequences in time $h(\|\formations\|,\maxdegree,\tw):=(f(\|\formations\|,\maxdegree)\cdot (\tw+1)+2)^{f(\|\formations\|,\maxdegree)\cdot g(\|\formations\|,\maxdegree)\cdot (\tw+1)}$, and exclude any sequence that violates Condition~\ref{cond:seq-trans-valid}.
\end{proof}

\subsection{$\FPT$ Algorithm}
\label{sec:fpt-algorithm}
In this section we present our $\FPT$ algorithm. We first introduce the table data structure that is computed for each bag.

\paragraph{Data Structure} For each bag $j\in J$, create a table $\table_j$ of size $|\poss{j}|$. Each row $\ell$ in the table corresponds to a candidate condensed sequence, and consists of three entries:
\begin{enumerate}
    \item $\pattern^j_\ell \in \poss{j}$ - the $\ell$\textsuperscript{th} condensed sequence on $j$.
    \item $\cost^j_\ell \in \NN \cup \{\infty\}$ - the (encountered) minimal number of configurations to cover all of $V_\downarrow (j)$ with a traversal $\calX$ such that $\cproject{\calX}{j}=\pattern^j_\ell$. Initialized to $\infty$.
    \item $\pointers^j_\ell$ - pointers to entries in the tables of the (one or two) children of $j\in J$. Initialized to \Null.
\end{enumerate}
Some rows are then deleted from the following tables:
\begin{itemize}
    \item For the root $\treeroot$ of $\calT$, all rows that contain a pattern with an $\uparrow$ symbol are deleted.
    \item If $\xx_0 \in \configs_{B_j}$, keep in $\table_j$ only rows with patterns that start with configuration $\xx_0$.
    \item If $\xx_f \in \configs_{B_j}$, keep in $\table_j$ only rows with patterns that end with configuration $\xx_f$.
\end{itemize}

Next, we define \emph{reduce}, \emph{lift} and \emph{combine}. These definitions will help express how a signature of a traversal $\calX$ at an add, forget and join nodes $j\in J$ respectively, are related to the signature of their children bag(s).

\paragraph{Reduce} For a condensed sequence $\calX=(\xx^i)_i \in \configs\cup\{\uparrow,\downarrow\}$, and a set of vertices $A \subseteq V$, let $\reduce{\calX}{A}$ be the sequence obtained from $\calX$ by first changing to $\uparrow$ any entry $\xx^i$ of $\calX$ such that $\xx^i \in \configs$ and $\activated{\xx^i} \cap A = \emptyset$ and then condensing the resultant sequence. Given traversal $\calX$ and $j$ is an add node, we have $\cproject{\calX}{j'}=\reduce{\cproject{\calX}{j}}{B_{j'}}$.

\paragraph{Lift} Similarly, $\lift{\calX}{A}$ is the condensed sequence obtained from changing to $\downarrow$ any $\xx^i \in \configs$ where $\activated{\xx^i} \cap A = \emptyset$. If $j \in J$ is a forget node, $\cproject{\calX}{j}=\lift{\cproject{\calX}{j'}}{B_j}$.

\paragraph{Combinations} For condensed sequences $\calX,\calY,\calZ$ of equal length, we say that $\calX$ \emph{combines} $\calY$ and $\calZ$ if for all $i$:
\begin{itemize}
    \item If $\xx^i\in\configs\cup\{\uparrow\}$ then $\xx^i=\yy^i=\zz^i$.
    \item If $\xx^i=\downarrow$ then either $(\yy^i,\zz^i)=(\downarrow,\uparrow)$ or $(\yy^i,\zz^i)=(\uparrow,\downarrow)$.
\end{itemize}
Indeed, for a given traversal $\calX$ and a join node $j\in J$ with children $j',j''\in J$, $\cproject{\calX}{j}$ combines $\cproject{\calX}{j'}$ and $\cproject{\calX}{j''}$.

\paragraph{The process} The tables for each $j\in J$ are scanned from the leaves in $\calT$ up as specified in \textsf{UpdateAllTables} (Algorithm~\ref{algorithm:update-all-tables}). The leaves in $\calT$ consist of empty bags, and therefore the only possible condensed sequence is $(\uparrow)$ for which the cost is set to $0$. The subroutine $\textsf{UpdateTable}$ is then used to update parent nodes in $\calT$ until reaching the root. \textsf{UpdateTable} considers each node type of the nice tree decomposition: \emph{add}, \emph{forget} and \emph{join}, and handles it accordingly.

\begin{algorithm}[ht]
\caption{\textsf{UpdateAllTables}}\label{algorithm:update-all-tables}
\begin{algorithmic}

    \STATE {\bf for} each bag $j \in J$, from leaves up {\bf do} \textsf{UpdateTable}(j);
    
    \hrulefill

    \STATE \underline{\textsf{UpdateTable}($j$)}:

    \FOR{$\pattern^j_\ell \in \textsf{enumerate\_patterns}(j)$}
        \IF{$j$ is a \codebox{leaf} node in $\calT$}
            \STATE {\bf if} $\pattern^j_\ell=(\uparrow)$ {\bf then} set $\cost^j_\ell=0$;
        \ELSIF{$j$ is an \codebox{add} node, with $B_j=B_{j'}\cup\{v\}$, where $j'$ is the child of $j$ in $\calT$}
            \STATE Let $\ell'$ in $\table_{j'}$ satisfy $\pattern^{j'}_{\ell'}=\reduce{\pattern^j_\ell}{B_{j'}}$;
            \STATE {\bf if} $\pattern^j_\ell$ visits $v$ {\bf then} set $\cost^j_\ell=\cost^{j'}_{\ell'}$;
            \STATE Set $\pointers^j_\ell:=\ell'$;
        \ELSIF{$j$ is a \codebox{forget} node, with $B_j=B_{j'}\setminus\{v\}$, where $j'$ is the child of $j$ in $\calT$}
            \STATE Let $L'=\{\ell': \pattern^j_\ell = \lift{\pattern^{j'}_{\ell'}}{B_j}\}$.
            \STATE Set $\ell' = \argmin_{\ell'\in L'}{\cost^{j'}_{\ell'}+|\{i: (\pattern^{j'}_{\ell'})^i\cap B_j = \emptyset\}|}$;
            \STATE Set $\cost^j_\ell=\cost^{j'}_{\ell'}+|\{i: (\pattern^{j'}_{\ell'})^i\cap B_j = \emptyset\}|$;
            \STATE Set $\pointers^j_\ell:=\ell'$;
        \ELSIF{$j$ is a \codebox{join} node, with childs $j',j''$ in $\calT$}
            \STATE Let $L^2=\{(\ell',\ell''): \pattern^j_\ell$ ~\text{combines}~ $\pattern^{j'}_{\ell'}$ ~\text{and}~  $\pattern^{j''}_{\ell''}\}$;
            \STATE Set $(\ell',\ell'')=\argmin_{(\ell',\ell'')\in L^2}{\cost^{j'}_{\ell'} + \cost^{j''}_{\ell''}}$;
            \STATE Set $\cost^j_\ell=\cost^{j'}_{\ell'} + \cost^{j''}_{\ell''}$;
            \STATE Set $\pointers^j_\ell=(\ell',\ell'')$;
        \ENDIF
    \ENDFOR
\end{algorithmic}
\end{algorithm}

After updating $\table_\treeroot$ where $\treeroot\in J$ is the root in $\calT$, let $\ell$ be the row with lowest cost. By following the pointers, we can \emph{reconstruct} an optimal traversal $\calX$. The detailed proof is provided in Appendix~\ref{sec:omitted-proofs}.

\begin{restatable}{thm}{fpt}
\label{theorem:fpt}
$\MRFGC$ can be solved in time $\calO(n \cdot h(\|\formations\|,\maxdegree,\tw)$, $\FPT$ in $\|\formations\|,\maxdegree,\tw$. In particular, $\MRCGC$ is $\FPT$ in $k,\maxdegree,\tw$.
\end{restatable}

Essentially, Theorem~\ref{theorem:fpt} follows from the following two observations. First, if the set of active vertices $\activated{\xx}$ intersects two bags $j',j''$, then it must intersect their common parent $j$.
Indeed, by the definition of tree decomposition, removing bag $j$ breaks the graph into disconnected components that separate bag $j'$ from bag $j''$. By the definition of a formation, the set of activated vertices is a connected sub-graph of $G$. Therefore, it must intersect bag $j$. This observation ensures that the costs are updated correctly, and no configuration is double-counted. Second, if two traversals $\calX,\calY$ have the same signature at $j$, we can replace $\reduce{\calX}{V_\downarrow (j)}$ with $\reduce{\calY}{V_\downarrow (j)}$, and get a valid traversal. Therefore, picking child signatures with minimal cost ensures an optimal traversal, breaking ties arbitrarily.

Although Theorem~\ref{theorem:fpt} is interesting theoretically, the complexity grows quickly with $\|\formations\|$, $\maxdegree$ and $\tw$, as we enumerate over configuration sequences. A natural follow-up question is whether efficient approximation algorithms exist. We next focus on trees ($\tw=1$), and show that computation time may be independent of $\maxdegree$, for a variety of formation families.

\section{Approximating $\MRFTC$ in $\PTAS$-time}
In this section, we study \emph{polynomial time approximation schemes ($\PTAS$)} for trees, that is, $\MRFTC$. A $\PTAS$ algorithm is given as input an error parameter $\varepsilon$, and should output a traversal $\calX_\approx$ that takes $\time{\calX_\approx} \le \timeoptimal\cdot (1+\varepsilon\cdot \error(k,\maxdegree))$ and runs in time $\poly(n) \cdot h(\frac{1}{\varepsilon},k)$. We stress that the run-time of the approximation algorithm is independent of the maximal degree $\maxdegree$ of the tree $T$. For $k=3$ connected robots, we show in Section~\ref{sec:approx-three-robots} that the approximation error is independent of $\maxdegree$ as well. For ease of exposition, we assume that $\xx_0=\xx_f=\treeroot$, the root of the input tree $T$.

We observe that if $(\formations,\transpositions)$ is \emph{collapsible}, an approximation of an optimal traversal can be computed in time independent of $\maxdegree$. Essentially, a formation is collapsible if the robots are always allowed to get closer. Recall that a \emph{contraction} of a graph $G$ along an edge $e=\{u,v\}$ is a graph $G'$ where $u,v$ are replaced with a single vertex $w$, and every edge in $G$ that was incident to either $u$ or $v$ is now incident to $w$ in $G'$.



\begin{Definition}
We say that $(\formations, \transpositions)$ is \emph{collapsible} if for each formation $\langle G_\alpha,\xx_\alpha \rangle \in \formations$ and each \emph{contraction} $G_{\alpha'}$ of $G_\alpha$, we have $\langle G_\alpha, \{\xx_\alpha, \xx_{\alpha'}\} \rangle \in \transpositions$. Configuration $\xx_\alpha'$ is defined by the graph contraction, where the number of robots of each type at the end-points of the contracted edge is added-up.
\end{Definition}

Intuitively, the idea is to cover the tree $T$ with a collection of $\calO(n\varepsilon)$ sub-trees of size $\calO(1/{\varepsilon})$, for which an optimal traversal can be found with an exhaustive search. Then, the approximate traversal is defined to traverse each such tree optimally, and spend $\calO(\|\formations\|)$ time to re-group at the root of each sub-tree. It is possible for the robots to re-group at some occupied node in $|V_\alpha-1|$ steps, since we assume $(\formations,\transpositions)$ are collapsible. Recall that $\timeoptimal$ is the optimal time to cover the tree, denote by $\timecover$ the sum of optimal times to cover each $\calO(1/{\varepsilon})$ sub-tree in the cover, and $\timegreedy$ the time it takes to cover the tree with the greedy algorithm. Then, we have three tasks at hand:
\begin{enumerate}
    \item Efficiently find such an $\varepsilon$ tree-coverage.
    \item Bound from above $\timegreedy - \timecover = \calO(f_+(\|\formations\|) n\varepsilon)$.
    \item Bound from below $\timecover - \timeoptimal = \calO(f_-(\|\formations\|,\maxdegree) n\varepsilon)$.
\end{enumerate}
We start by defining a tree-cover:

\begin{Definition}
A \emph{tree-cover} $\calP=(\calT,\calV)$ of $T$ is a tree $\calT$ and a family $\calV=(V_\tau)_{\tau \in V(\calT)}$ of sub-trees of $V$ such that:
\begin{enumerate}
    \item \textbf{Coverage}: It covers $V$, that is, $\bigcup_{\tau \in V(\calT)}{V_\tau}=V$.
    \item \textbf{Small Overlap}: For any $\tau \neq \tau' \in V(\calT)$, $|V_\tau \cap V_{\tau'}| \le 1$.
    \item \textbf{Connectivity}: For any $\tau \neq \tau' \in V(\calT)$, $\{\tau,\tau'\} \in E(\calT)$ iff $|V_\tau \cap V_{\tau'}|=1$, and the common vertex $v$ is a leaf in $\tau$ and a root in $\tau'$, or vice versa, or it is the common root.
\end{enumerate}
\end{Definition}

Note that $\calV$ uniquely identifies the tree-cover, and therefore we can denote it by $\calP(\calV)$. Next, we parameterize a tree-coverage with the maximal size of a sub-tree in the coverage:

\begin{Definition}
Let $T=(V,E,\treeroot)$ be a rooted tree and $\varepsilon > 0$. An $\varepsilon$-tree-cover $\calP=(\calT,\calV)$ is a tree-cover where each sub-tree $\tau \in \calT$ satisfies $|\tau| \le \frac{2}{\varepsilon}$, and $|\calV| \le n\varepsilon+1$.
\end{Definition}

\textsf{TreeCover} (Algorithm~\ref{algorithm:cover-t}) efficiently computes an $\varepsilon$-tree-cover. In particular, $\varepsilon$-tree-cover always exist:

\begin{algorithm}[ht]
\caption{\textsf{TreeCover}}\label{algorithm:cover-t}
\begin{algorithmic}
    \REQUIRE A rooted tree $T=(V,E,\treeroot)$;
    \bindent
    \STATE ~ Parameter $0 < \varepsilon < 1$;
    \eindent
    \ENSURE The size $0\le \textsf{size} < \frac{1}{\varepsilon}$ of remaining tree to be covered;
    \bindent
    \STATE ~~~\, The sub-tree $\tau$ remained to be covered;
    \STATE ~~~\, A family $\calV$ of sub-trees of $V \setminus \tau$, with $|\calV| \le n\varepsilon$, where the size of each tree is in $[\frac{1}{\varepsilon}, \frac{2}{\varepsilon}]$;
    \STATE ~~~\, $\configs(\calV \cup \{\tau\})$ is an $\varepsilon$-tree-cover of $T$;
    \eindent
    
    \hrulefill

    \STATE Initialize $\textsf{size}\gets 1$, $\tau \gets \textsf{Tree}(\{\treeroot\})$, $\calV \gets \emptyset$;
    
    \IF{$\treeroot$ is a leaf}
        \RETURN $\textsf{size}, \tau, \calV$;
    \ENDIF
    \FOR{$u \in \children{\treeroot}$}
        \STATE $\textsf{size}', \tau', \calV' \gets \textsf{TreeCover}((V,E,u),\varepsilon)$;
        \STATE $\textsf{size}$ += $\textsf{size}'$;
        \STATE $\tau.\textsf{add\_subtree}.(\tau')$;
        \STATE $\calV$.\textsf{add}($\calV'$);
        \IF{$\textsf{size} > \frac{1}{\varepsilon}$}
            \STATE $\textsf{size} \gets 1$;
            \STATE $\calV$.\textsf{add}($\{\tau\}$);
            \STATE $\tau \gets \textsf{Tree}(\{\treeroot\})$;
    \ENDIF
    \ENDFOR
    \RETURN $\calV, \tau, \textsf{size}$;
    
\end{algorithmic}
\end{algorithm}

\begin{Lemma}
\label{lemma:alg-cover-correctness}
\textsf{TreeCover} runs in time $\calO(n \log(1/{\varepsilon}))$ and returns a tree-cover $\calP(\calV \cupdot \{\tau\})$ of $T$.
\end{Lemma}
\begin{proof}
As for complexity, \textsf{TreeCover} is a DFS search over $T$ starting from the root $\treeroot$. After each visit, basic operations such as adding (a pointer to) a sub-tree, adding (pointers / indices) of sub-trees to the cover, and adding-up / comparing $\calO(\log(\frac{1}{\varepsilon}))$-bit numbers.
\paragraph{Coverage.} Since DFS will scan the whole tree, eventually all vertices will appear in a sub-tree in $\calV \cup \{\tau\}$.
\paragraph{Small Overlap.} Once a sub-tree is added to $\calP$, all vertices except the root are forgotten, and therefore the intersection can only include the root, which is the current vertex that the DFS algorithm visits.
\paragraph{Connectivity.} In $\configs$, connectivity holds by definition, however, we must show that $\calT$ is indeed a tree. By keeping the root of each tree $\tau'$ that is added to $\calV$, a path between $\tau'$ and the root $\tau$ in $\calT$ is ensured by induction. Hence, $\calT$ is connected. In addition, it contains no cycles as it will translate to a cycle in $T$.
\end{proof}

In addition, \textsf{TreeCover} outputs an $\varepsilon$-tree-cover:

\begin{Lemma}
\label{lemma:alg-eps-cover-correctness}
\textsf{TreeCover} runs in time $\calO(n \log(\frac{1}{\varepsilon}))$ and returns an $\varepsilon$-tree-cover $\calP(\calV \cupdot \{\tau\})$ of $T$.
\end{Lemma}

\begin{proof}

By Lemma~\ref{lemma:alg-cover-correctness}, $\calP$ is a tree-cover. The size of each added tree always coincides with $\textsf{size}$ by induction. For the leaves it is $1$, and then as long as it doesn't pass $\frac{1}{\varepsilon}$, it is updated by adding the sizes of the sub-trees of each child (which are correct by induction). Once a tree is added to $\calV$, all vertices except the root are forgotten and the size of the tree is reset to $1$.

Therefore, since sub-trees are added to $\calV$ only after checking their size is greater than $\frac{1}{\varepsilon}$, we have that $|\tau| < \frac{1}{\varepsilon}$, as otherwise it would have been added to $\calV$.

Now, suppose in contradiction that there is a tree $T_p \in \calP$ rooted at $u$ of size greater than $\frac{2}{\varepsilon}$. Then there must be a child of $u$ for which \textsf{TreeCover} returned a tree of size $\ge \frac{1}{\varepsilon}$. But this is a contradiction, since we proved that the size of $\tau'$ is always strictly less than $\frac{1}{\varepsilon}$.

Finally, denote by $P := |\calV|$. Then $|V| = n = |\tau| + \sum_{\tau' \in \calV}{|\tau'|} \ge 0 + P \cdot \frac{1}{\varepsilon}$. Therefore, $P \le n\varepsilon$.
\end{proof}

\subsection{$\PTAS$ Algorithm}
Next, we describe \textsf{GreedyTraverse} (Algorithm~\ref{algorithm:greedy-traversal}). Given an $\varepsilon$ tree-coverage $\calP$ of $T$, it computes a traversal of the tree $T$ in time $\timeoptimal(1+n\varepsilon\cdot \textsf{error}_{\textsf{greedy}}(\|\formations\|,\maxdegree)$. The algorithm runs in time $n f_\textsf{greedy}(\varepsilon,\|\formations\|)$, independent of $\maxdegree$.

In order to bound $\textsf{error}_{\textsf{greedy}}$, we compare it with the total time to cover each sub-tree $\tau \in \calV$ individually, denoted $\timecover$, and show that the difference is bounded by $n\varepsilon \cdot \textsf{error}_{\textsf{greedy}}(\|\formations\|)$. We show that $\timecover\le \timeoptimal+n\varepsilon h_{\textsf{tree-cover}}(\|\formations\|,\maxdegree)$ and conclude that \textsf{GreedyTraverse} is a $\PTAS$ algorithm for $\MRCTC$.

\begin{algorithm}[ht]
\caption{\textsf{GreedyTraverse}}\label{algorithm:greedy-traversal}
\begin{algorithmic}
    \REQUIRE A rooted tree $(T,\treeroot)$, a collapsible $(\formations,\transpositions)$, a tree-cover $\calP$;
    \ENSURE A traversal $\calX_\approx$;
    
    \hrulefill

    \STATE Let $\calV_\treeroot \subseteq \calV$ be the sub-trees rooted at $\treeroot$;

    \FOR{sub-tree $\tau$ in $\calV_\treeroot$}
        \STATE $\calY_\tau \gets \MRFTC(\tau,\treeroot,k)$;
        \STATE Compute the following traversal $\calX_\tau$ of sub-tree $\subtree{\tau}$:
        \bindent
            \STATE Follow $\calY_\tau$;
            \IF{a leaf $v$ of $\tau$ is visited for the first time}
                \STATE Re-group at $v$;
                \STATE Follow $\textsf{GreedyTraverse}(\subtree{v},v,k,\calP|_{v})$;
                \STATE Get back to the configuration that visited $v$.
            \ENDIF
        \eindent
    \ENDFOR

    \STATE Set $\calX_\approx$ as the concatenation of the $\calX_\tau$'s;
    \RETURN $\calX_\approx$;
\end{algorithmic}
\end{algorithm}

Intuitively, \textsf{GreedyTraverse} traverses each sub-tree $\tau\in\calV$ in the tree-cover $\calP$ optimally. It can do so in time independent of $n$ as the size of each sub-tree of an $\varepsilon$ tree-cover is bounded by $2/\varepsilon$. In addition, whenever an optimal traversal of some sub-tree $\tau$ visits a leaf $v$ of $\tau$ for the first time, all the robots re-group at that leaf. Assuming the robots are in $\alpha$ form, they can re-group at $v$ in $|E_\alpha|$ steps, by contracting all edges in $G_\alpha$. Then, they traverse the sub-tree rooted at $v$ by recursively calling \textsf{GreedyTraverse}, and then return back to that configuration that visited $v$. Therefore, $f_+(\|\formations\|):=2\max_\alpha {|E_\alpha|}$. Note that the robots already start and end at $\treeroot$, so there is no need to account for $\tau_\treeroot$. Therefore, $\textsf{error}_\textsf{greedy}=f_+(\|\formations\|) n \varepsilon$.

Finally, we must bound $\timecover - \timeoptimal$. In an optimal traversal, robots may cover vertices from different sub-trees $\tau\neq\tau'\in\calV$ simultaneously, and therefore it could be that $\timecover > \timeoptimal$. In addition, the robots may get in-and-out of a sub-tree in a way that may help them cover the sub-tree faster. We observe that the latter cannot be the case for collapsible formations, since equivalently the robots may wait at the boundary (leaves of $\tau$ and its root) instead of leaving it. Hence, we show that the difference is bounded by $n\varepsilon f_-(\|\formations\|,\maxdegree)$.

\begin{Lemma}
\label{theorem:upper-bound}
$\timecover - \timeoptimal \le n\varepsilon f_-(\|\formations\|,\maxdegree)$.
\end{Lemma}
\begin{proof}
First, we prove that the time to cover a sub-tree $\tau$ is the same regardless of whether it is part of a bigger tree $T$ or not. We can view $\tau$ as a contraction of $T$ by contracting all the edges that are not in $\tau$. Then, any traversal in $T$ that covers $\tau$ is mapped by the contraction to a traversal within $\tau$ that covers $\tau$. The latter is a valid traversal of $\tau$ since $(\formations,\transpositions)$ is collapsible.

Next, consider the following traversal $\calX_\approx$. It follows the optimal traversal $\calX$, but every time the robots occupy the root of some sub-tree $\tau$ for the first time, the robots re-group at its root $\tau_\treeroot$, and then by enumerate over all $f(\|\formations\|,\maxdegree)$ possible configurations in $\configs_{\tau_\treeroot}$, by following the reverse process of re-grouping and then re-grouping at $\tau_\treeroot$. This takes at most $f_-(\|\formations\|,\maxdegree):= 2\max_\alpha |E_\alpha| \cdot f(\|\formations\|,\maxdegree)$ time. Clearly, $\time{\calX_\approx} \le \timeoptimal + f_-(\|\formations\|,\maxdegree)n\varepsilon$. On the other hand, we also have $\timecover \le \time{\calX_\approx}$.
\end{proof}
Therefore, we obtain the following result:
\begin{Theorem}
\label{theorem:ptas-mrftc}
Assume $(\formations,\transpositions)$ is collapsible. Then $\MRFTC$ is in $\PTAS$. Specifically, there exist an algorithm that computes a traversal of $T$ in time $\timeoptimal (1+n\varepsilon f_\approx(\|\formations\|,\maxdegree))$. The approximate traversal can be computed in time $\calO(n \cdot g_\approx(\|\formations\|,\varepsilon))$.
\end{Theorem}
\section{Approximate 3-Robot $\MRCTC$}
\label{sec:approx-three-robots}
\begin{figure}[htb]
\centering
\begin{subfigure}{0.24\linewidth}
    \includegraphics[width=\linewidth]{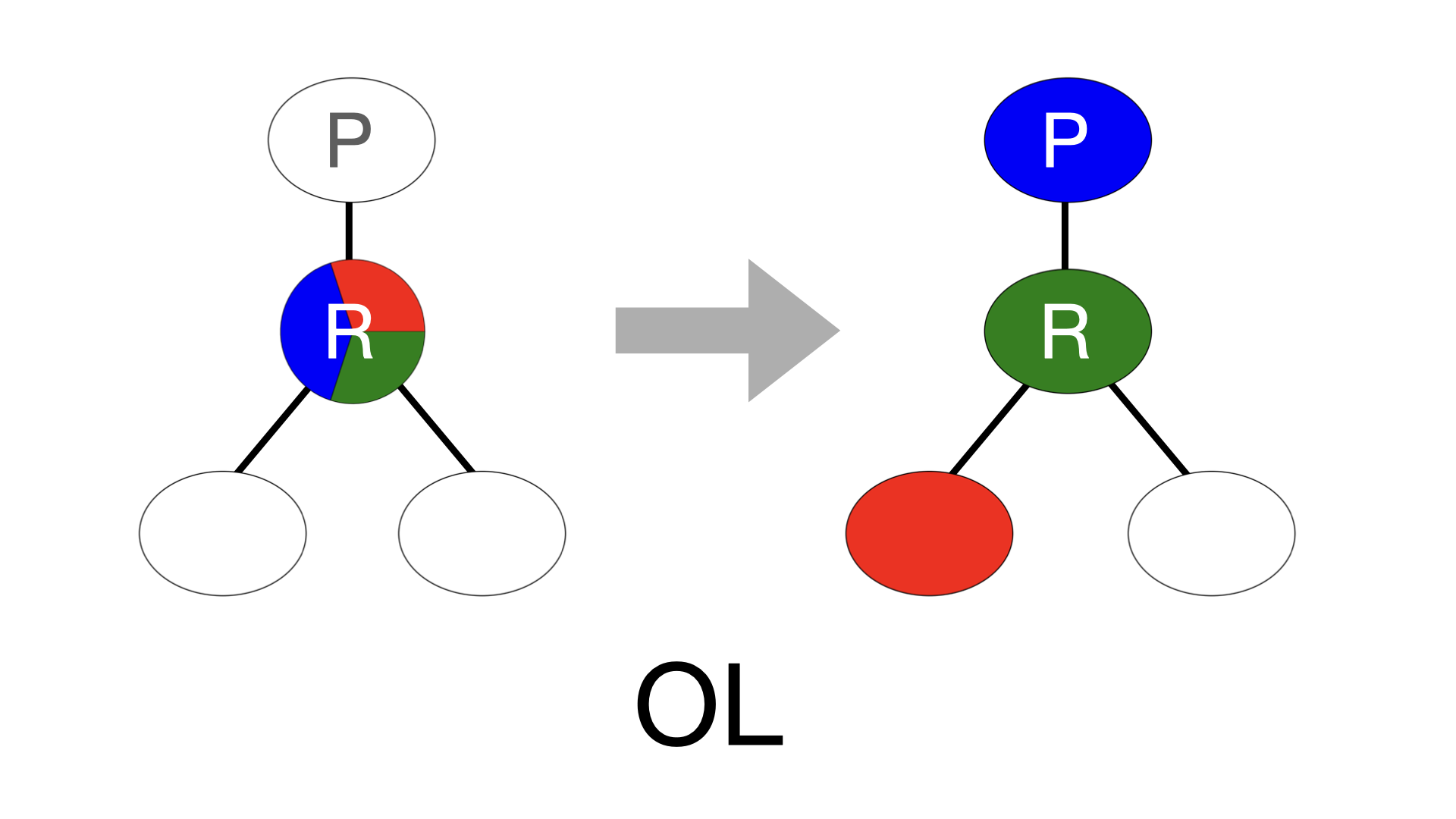}
\end{subfigure}
\begin{subfigure}{0.24\linewidth}
    \includegraphics[width=\linewidth]{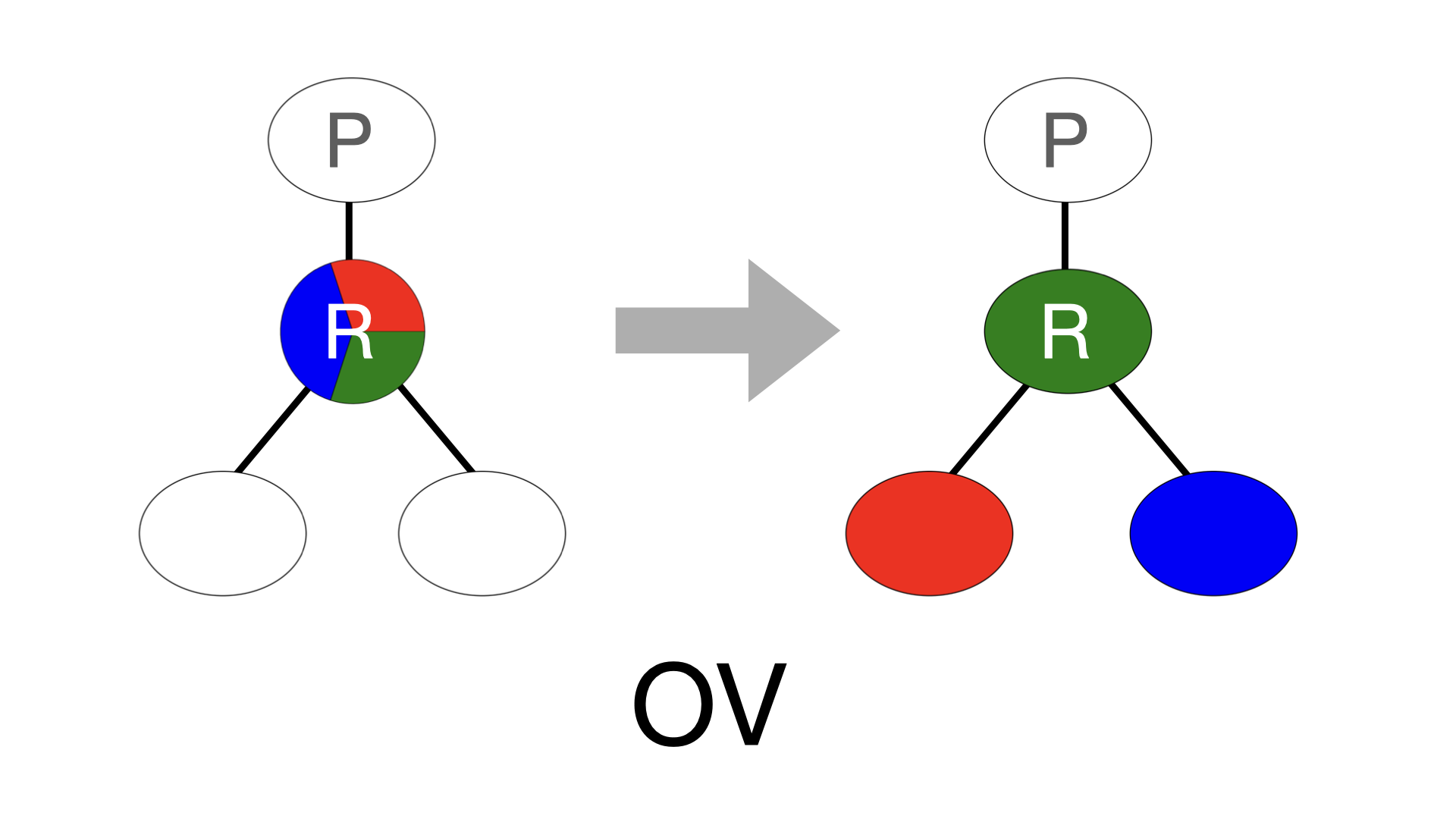}
\end{subfigure}
\begin{subfigure}{0.24\linewidth}
    \includegraphics[width=\linewidth]{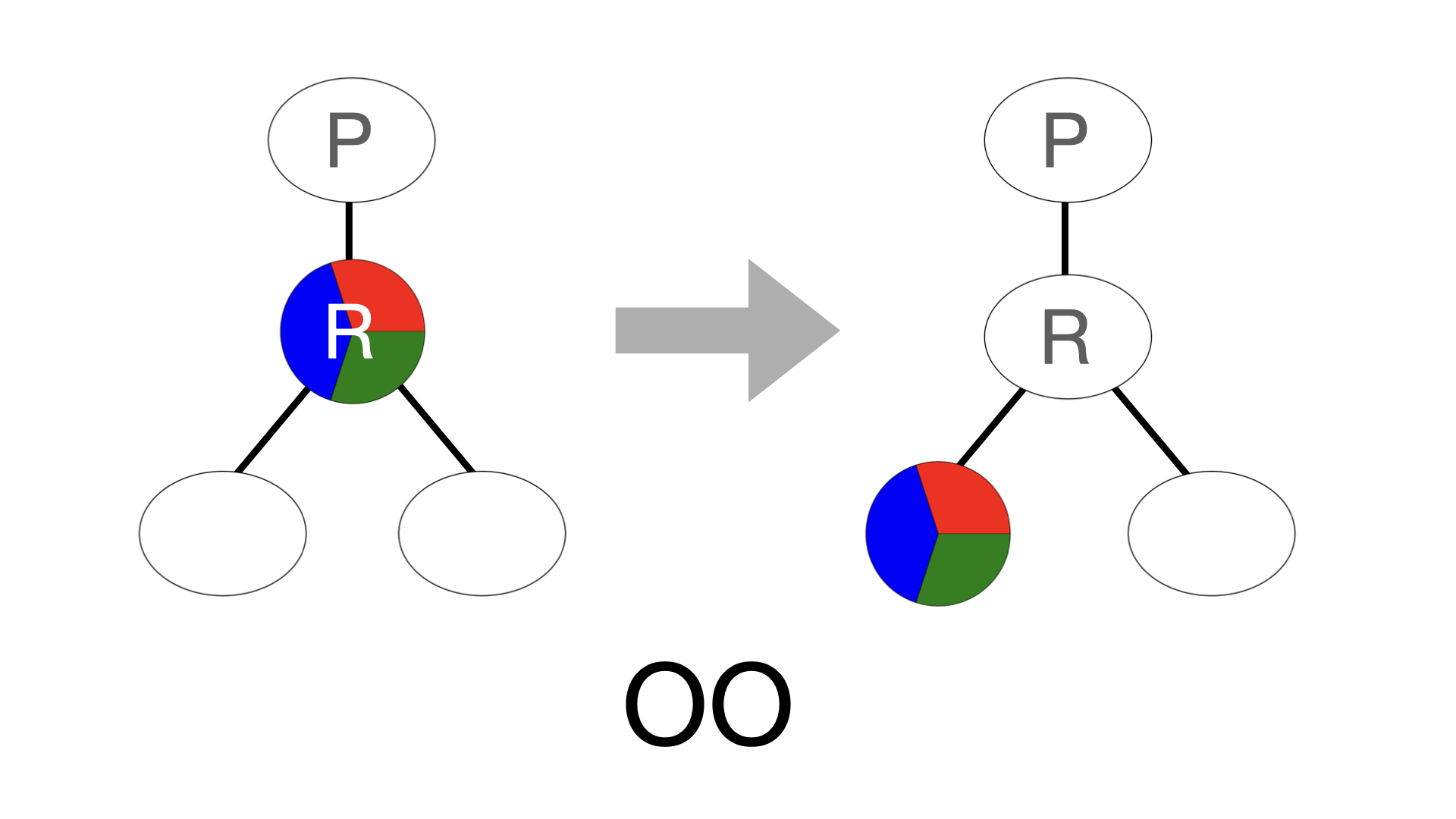}
\end{subfigure}
\begin{subfigure}{0.24\linewidth}
    \includegraphics[width=\linewidth]{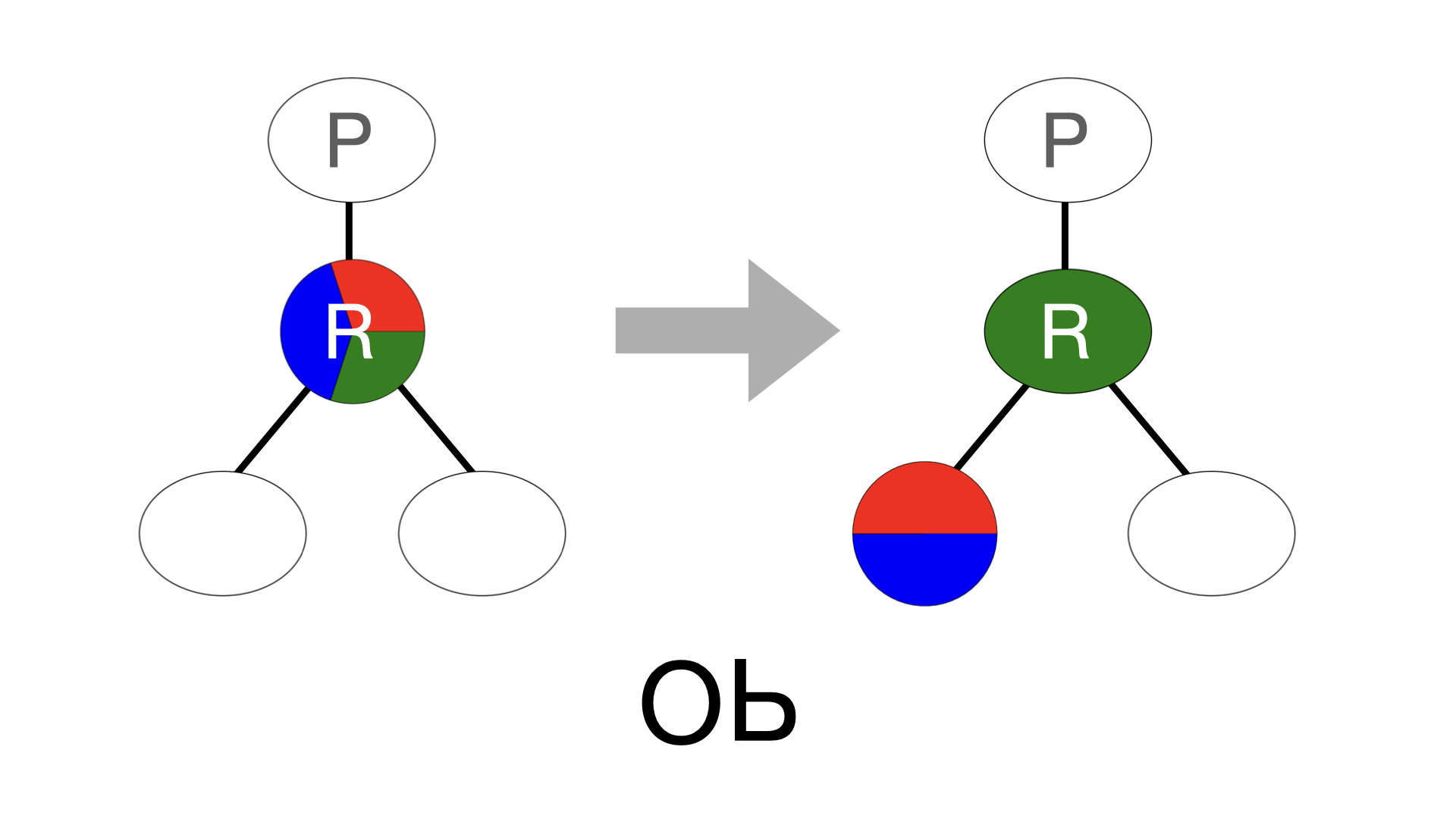}
\end{subfigure}

\begin{subfigure}{0.24\linewidth}
    \includegraphics[width=\linewidth]{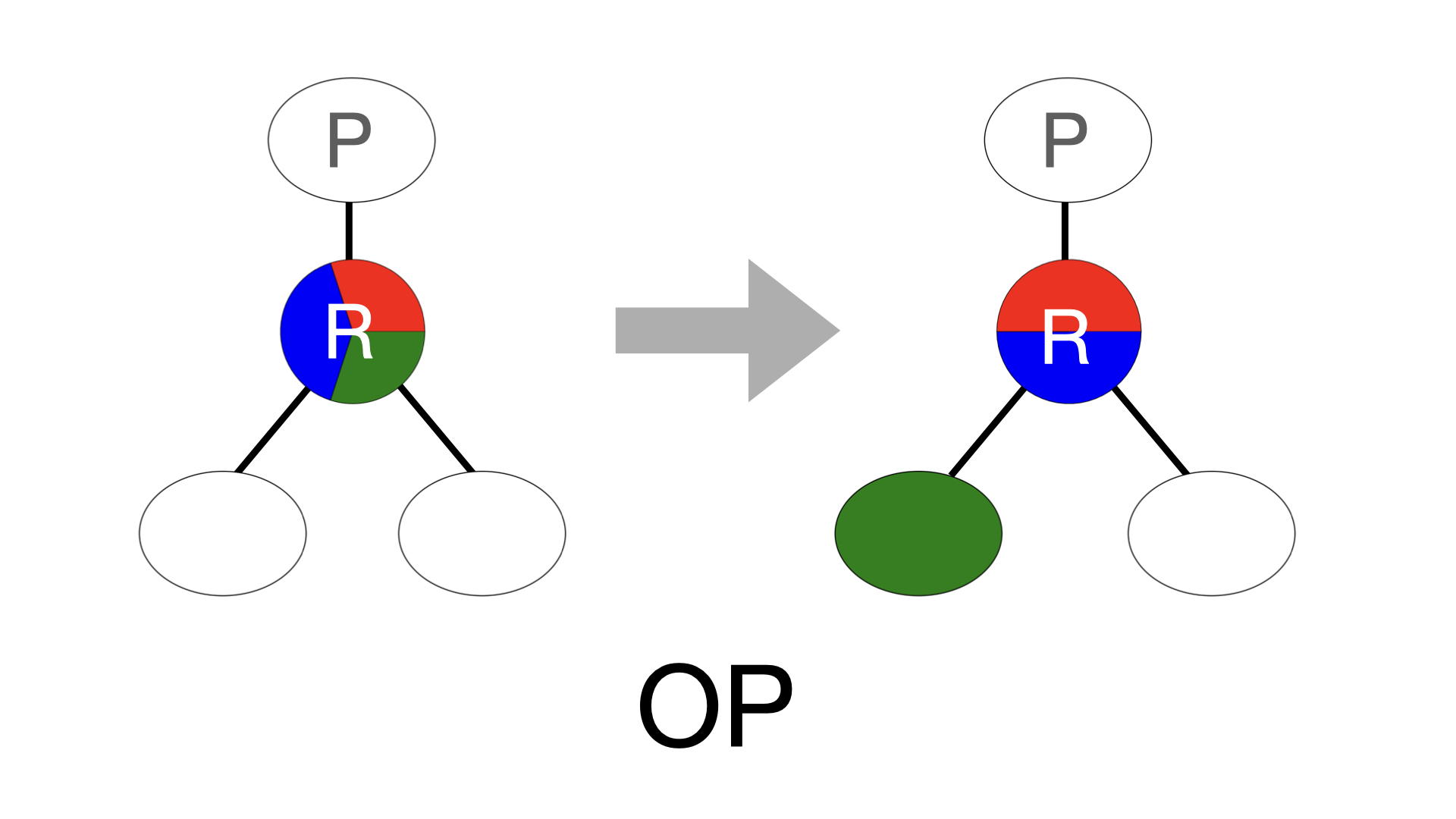}
\end{subfigure}
\begin{subfigure}{0.24\linewidth}
    \includegraphics[width=\linewidth]{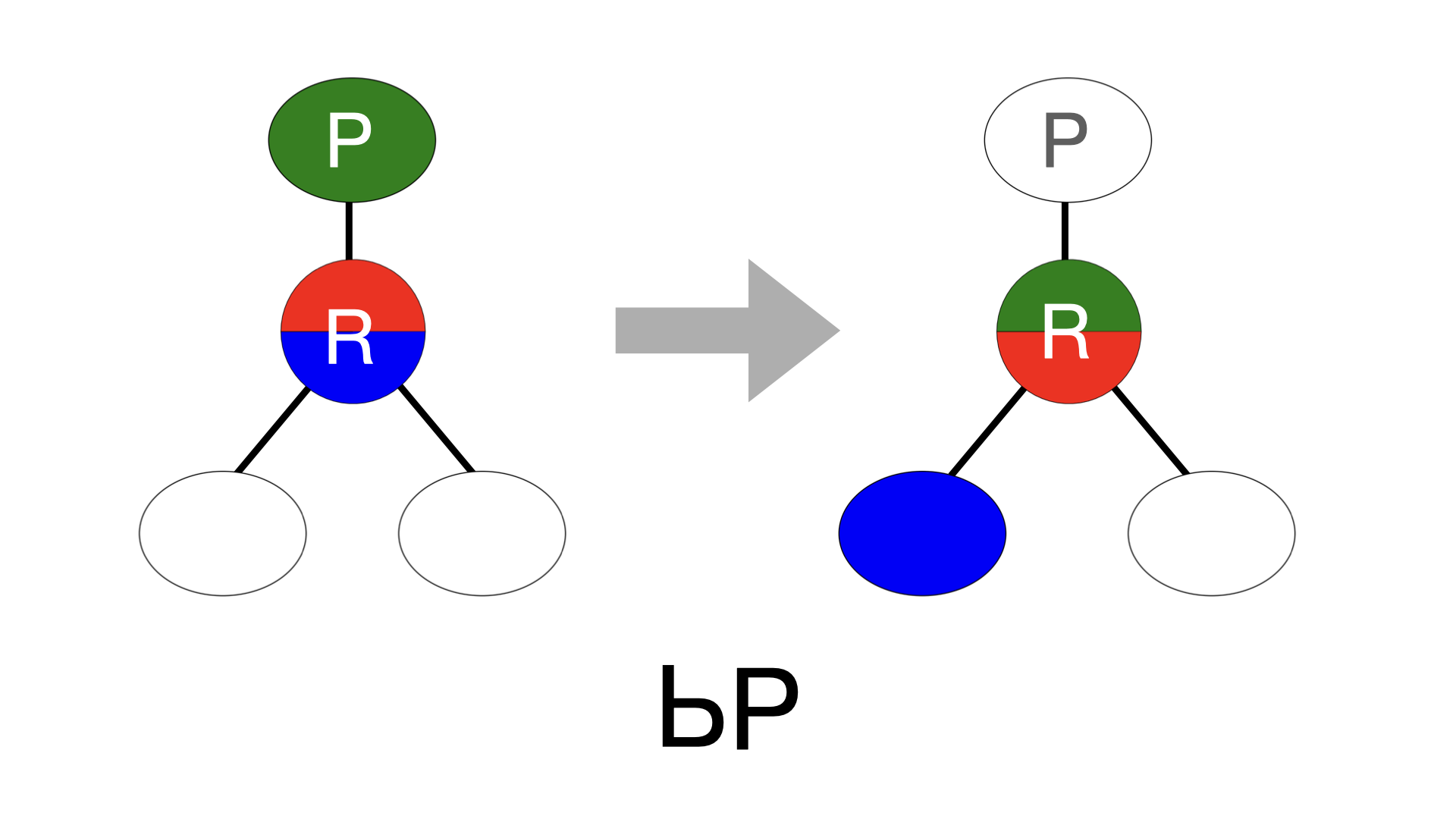}
\end{subfigure}
\begin{subfigure}{0.24\linewidth}
    \includegraphics[width=\linewidth]{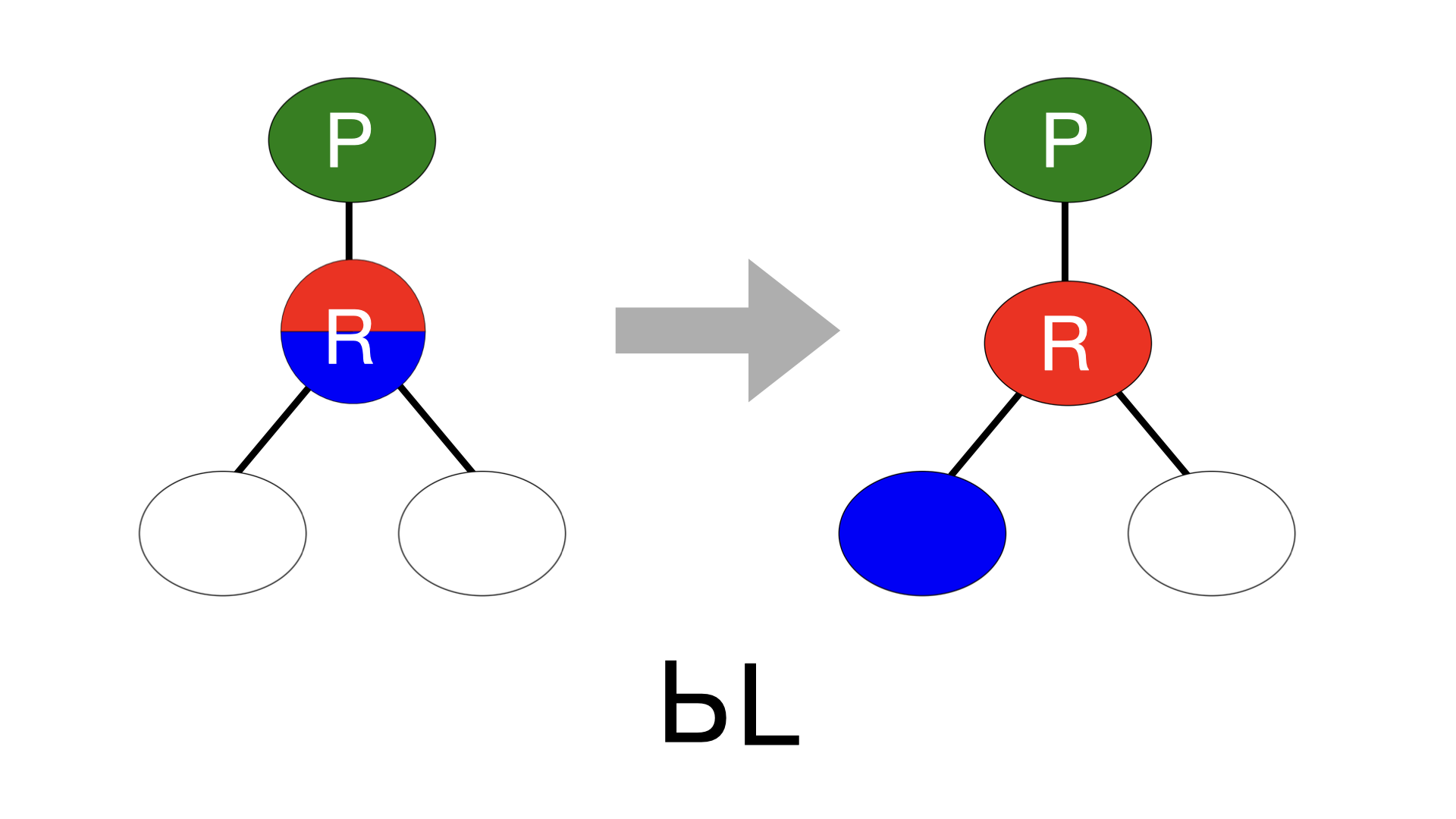}
\end{subfigure}
\begin{subfigure}{0.24\linewidth}
    \includegraphics[width=\linewidth]{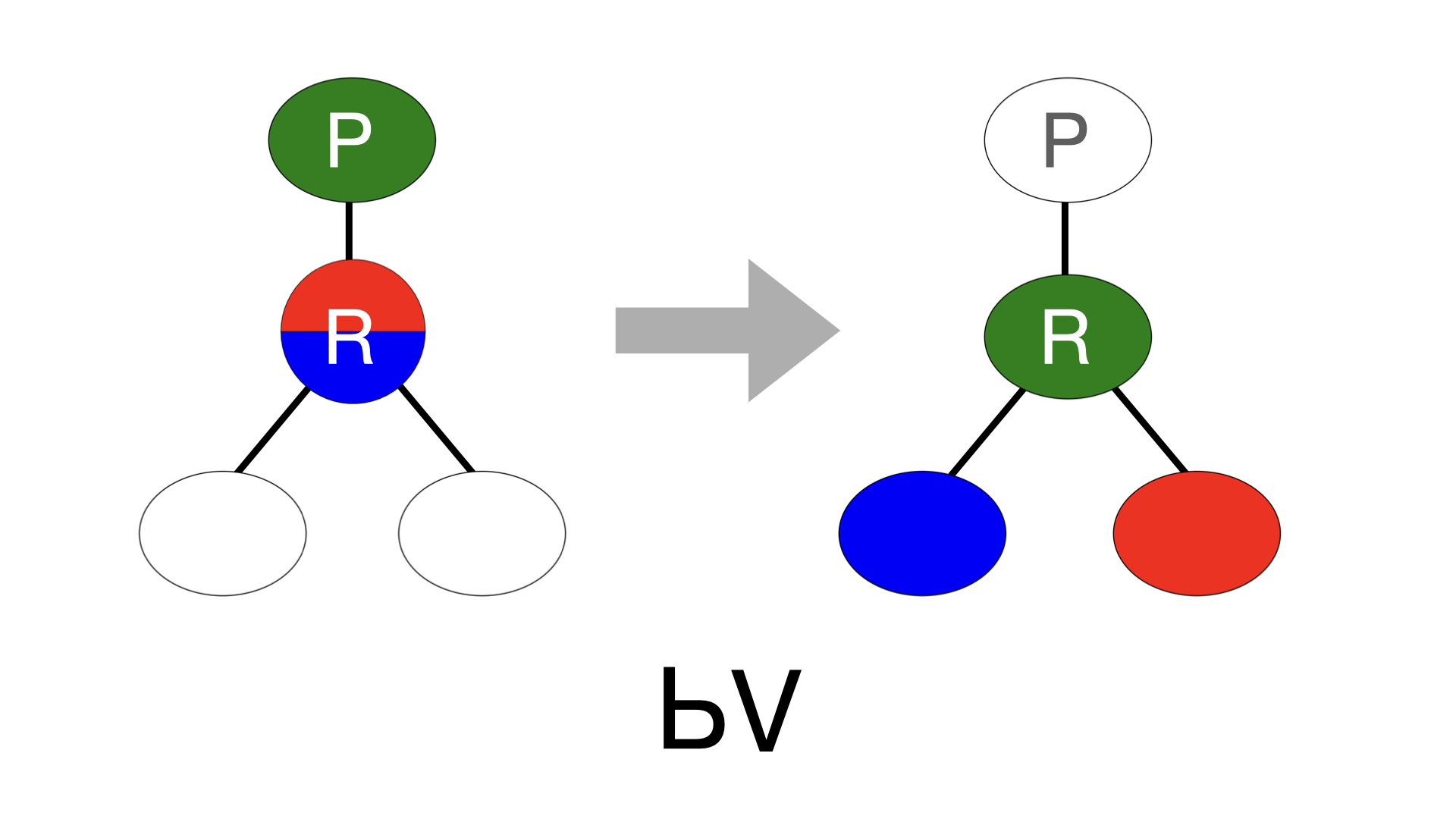}
\end{subfigure}

\begin{subfigure}{0.24\linewidth}
    \includegraphics[width=\linewidth]{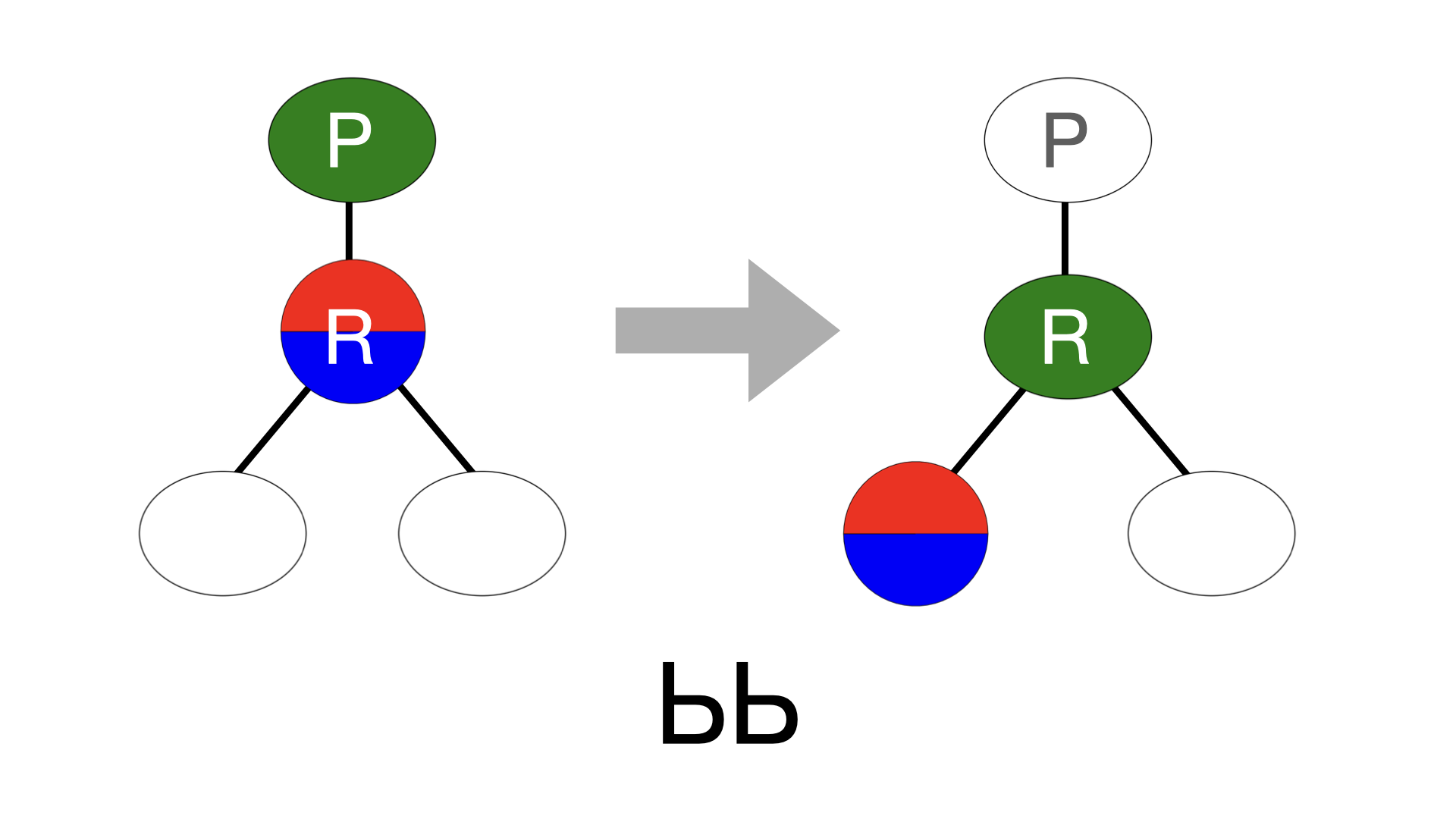}
\end{subfigure}
\begin{subfigure}{0.24\linewidth}
    \includegraphics[width=\linewidth]{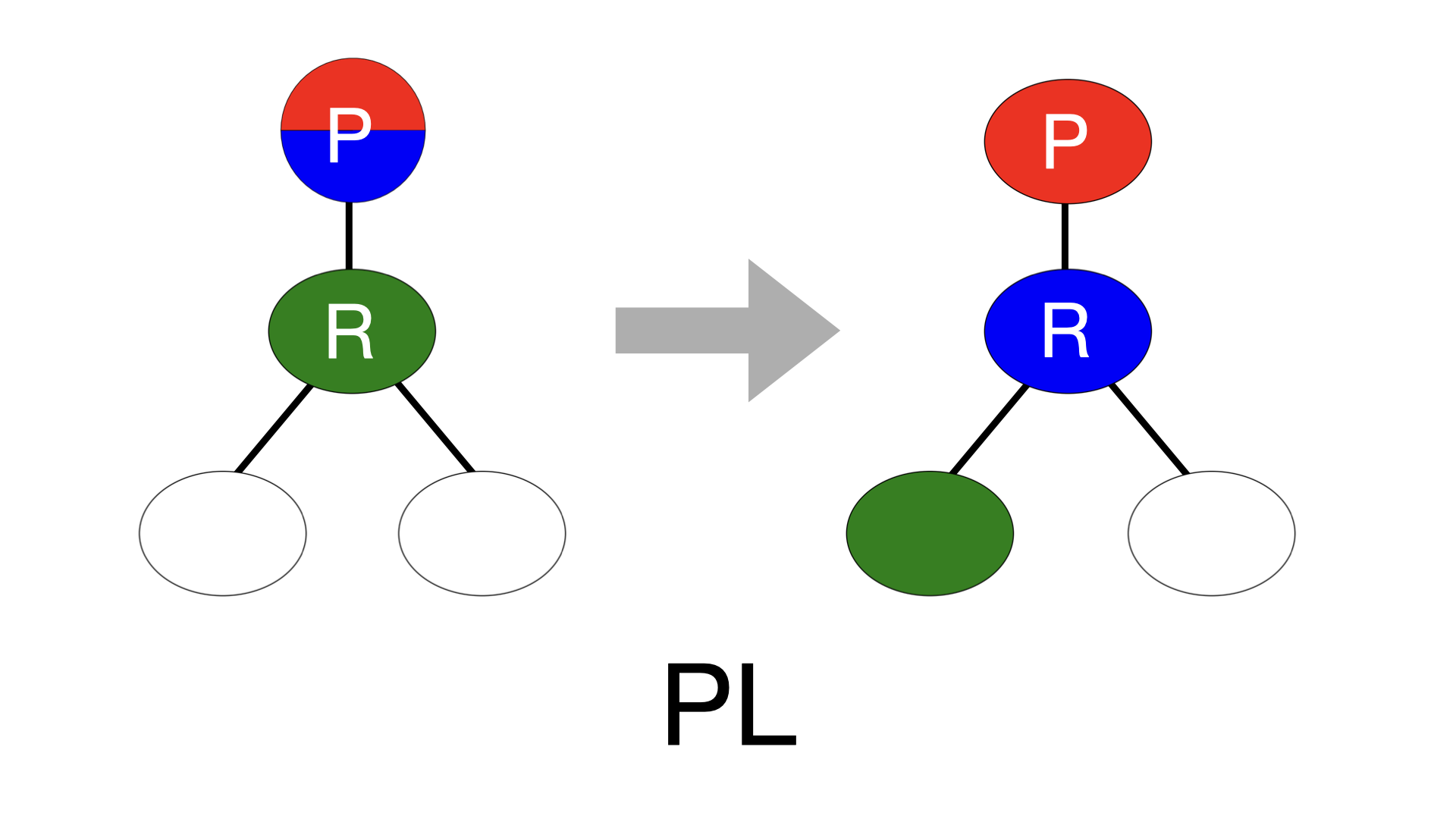}
\end{subfigure}
\begin{subfigure}{0.24\linewidth}
    \includegraphics[width=\linewidth]{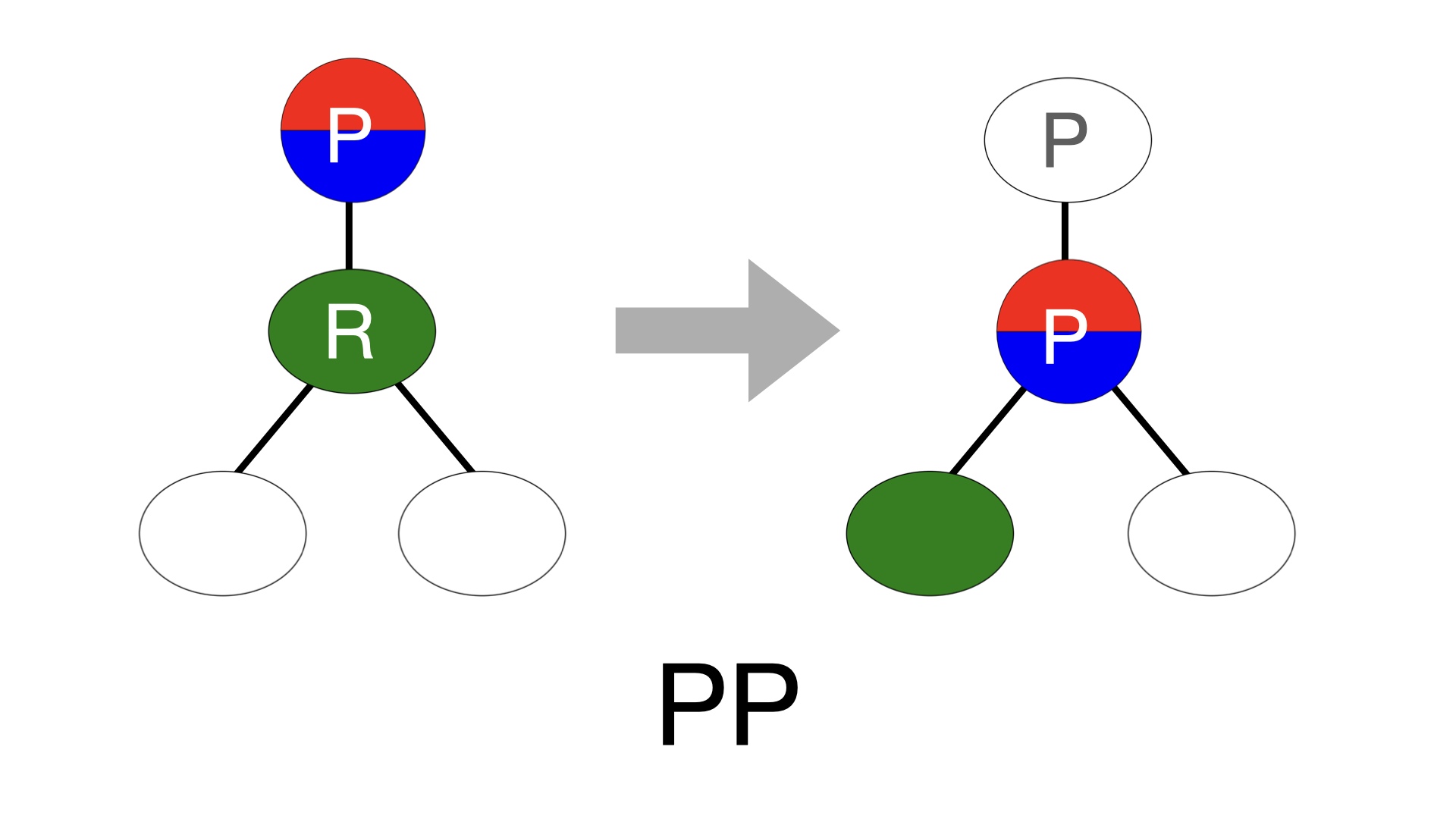}
\end{subfigure}
\begin{subfigure}{0.24\linewidth}
    \includegraphics[width=\linewidth]{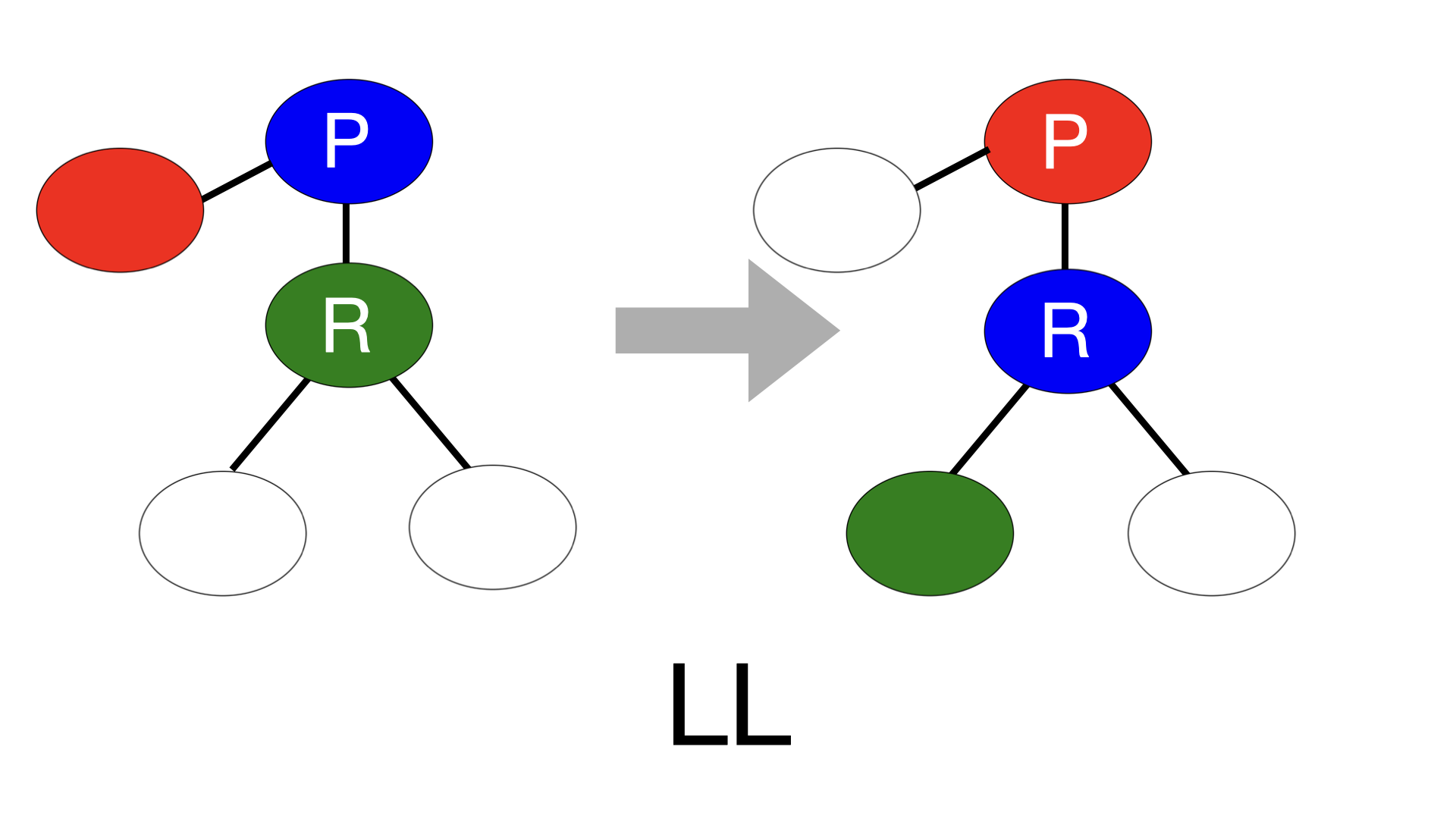}
\end{subfigure}

\caption{Transition Shapes for $k=3$ Robots} \label{fig:transition-families-3}
\end{figure}

For the special case of $\MRCTC$ with three robots, we prove that the approximation error of \textsf{GreedyTraverse} is independent of $\maxdegree$ as well.
\begin{Proposition}
\label{prop:upper-bound-k3}
For $k=3$, $\timecover - \timeoptimal \le 52 n\varepsilon$.
\end{Proposition}
Intuitively, we provide a tight analysis of transitions of $3$-robot configurations in trees. We classify the transitions into 12 categories, depicted in Figure~\ref{fig:transition-families-3}. We use a technique similar to the one used in the proof of the Z-Lemma~\ref{lemma:z-method}, to prove that there exist an optimal traversal where no transition category repeats, entering the same sub-tree. In such optimal traversal, the number of times that robots enter each sub-tree is bounded by 12, independent of $\maxdegree$. We may now construct a sub-optimal traversal that enters each sub-tree once by gluing all visits of a subtree with re-grouping at the tree-root. Since the number of re-groupings is bounded by 12, and the cost of re-grouping is $\le 2(k-1)=4$, this results with a traversal that visits each sub-tree separately, and hence its traversal time is $\ge \timecover$. Since $\timegreedy-\timecover \le 2(k-1)n\varepsilon$, we get an overall error that is $\le 52 n\varepsilon k$.

In Figure~\ref{fig:transition-families-3}, we enumerate over all possible \emph{transition shapes}, that start from a configuration that occupies some vertex $R$ with parent $P$, and end at a configuration that occupies at least one of the children of $R$. We use red, blue and green for the colors of the three robots, in order to demonstrate which robot goes where. A transition shape specifies how many robots visit $R$,$P$, a neighbor of $P$ (other than $R$), and a child of $R$, before and after the transition. We also distinguish between a configuration where the same child is occupied by 2 (\flip{P}) or 3 (O) robots, and a configuration where two children are occupied (V). An exhaustive search yields 12 such transition shapes. We want to show that any repetition of a transition shape can be avoided:

\begin{Lemma}
\label{lemma:z-method-shapes-3}
There exists an optimal \emph{tree} traversal with $k=3$ connected robots, where no transition shape repeats.
\end{Lemma}

\begin{figure}[htb]
\centering
\begin{subfigure}{0.35\linewidth}
    \includegraphics[width=\linewidth]{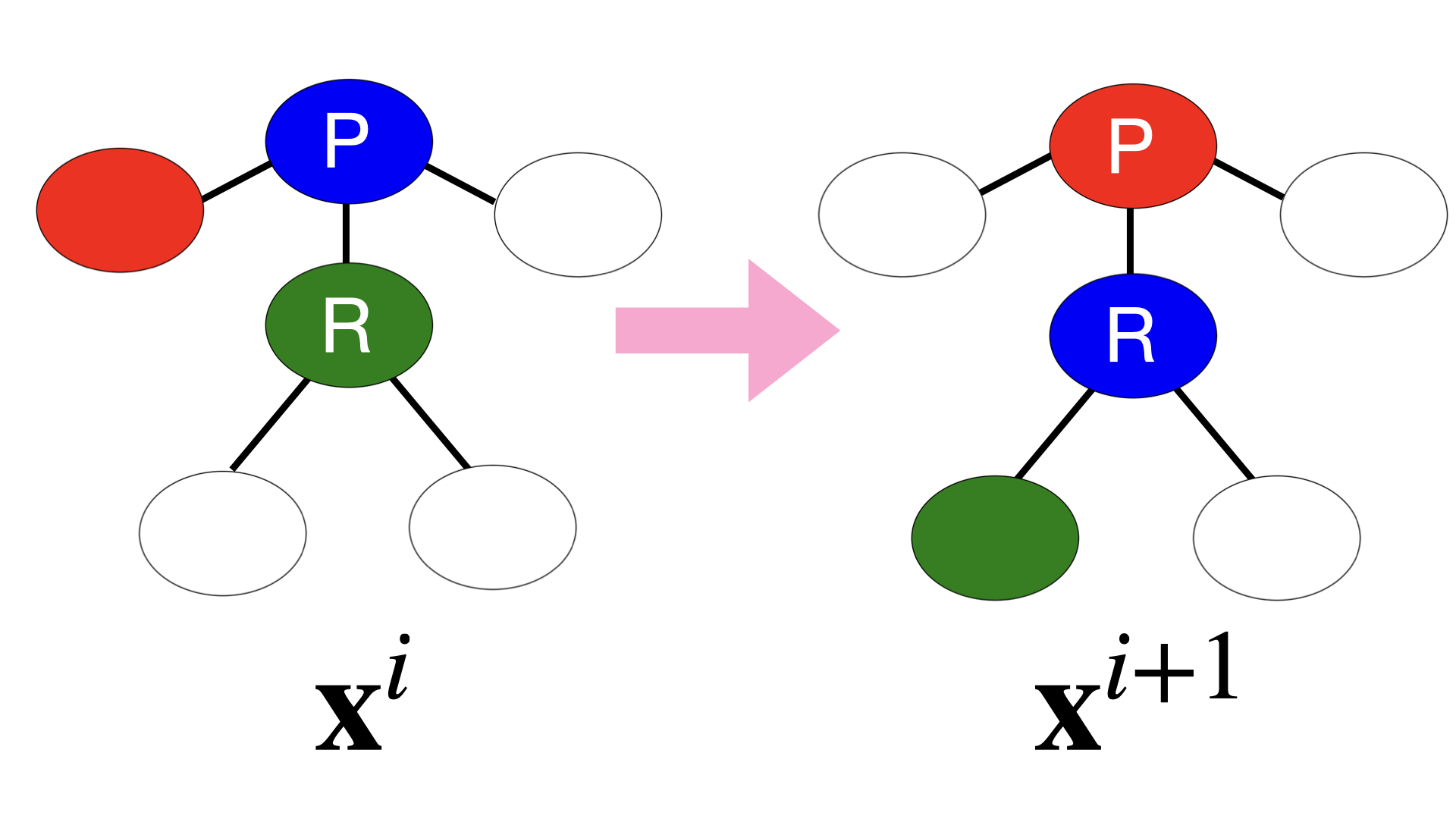}
\end{subfigure}
\begin{subfigure}{0.35\linewidth}
    \includegraphics[width=\linewidth]{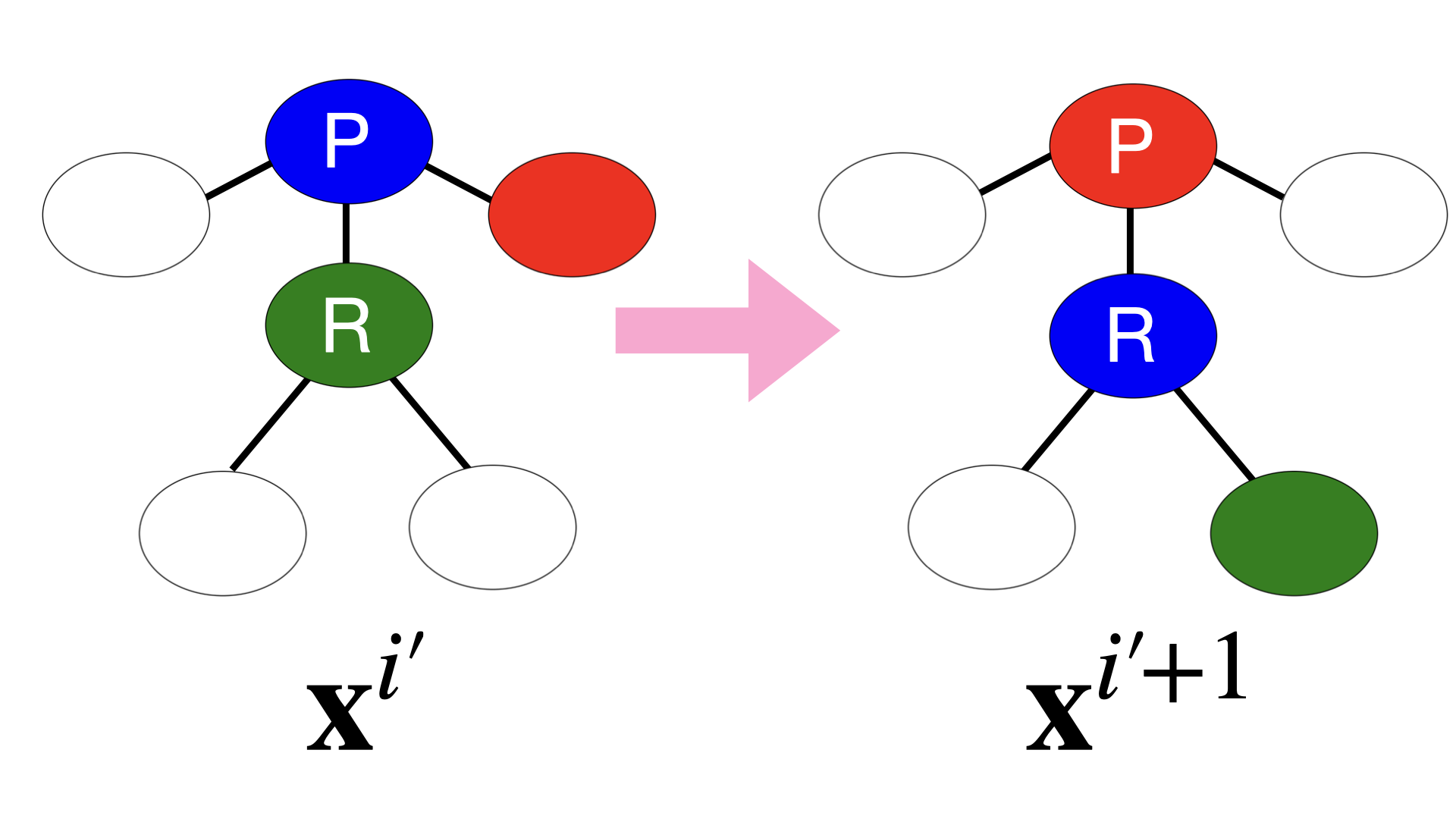}
\end{subfigure}

\begin{subfigure}{0.35\linewidth}
    \includegraphics[width=\linewidth]{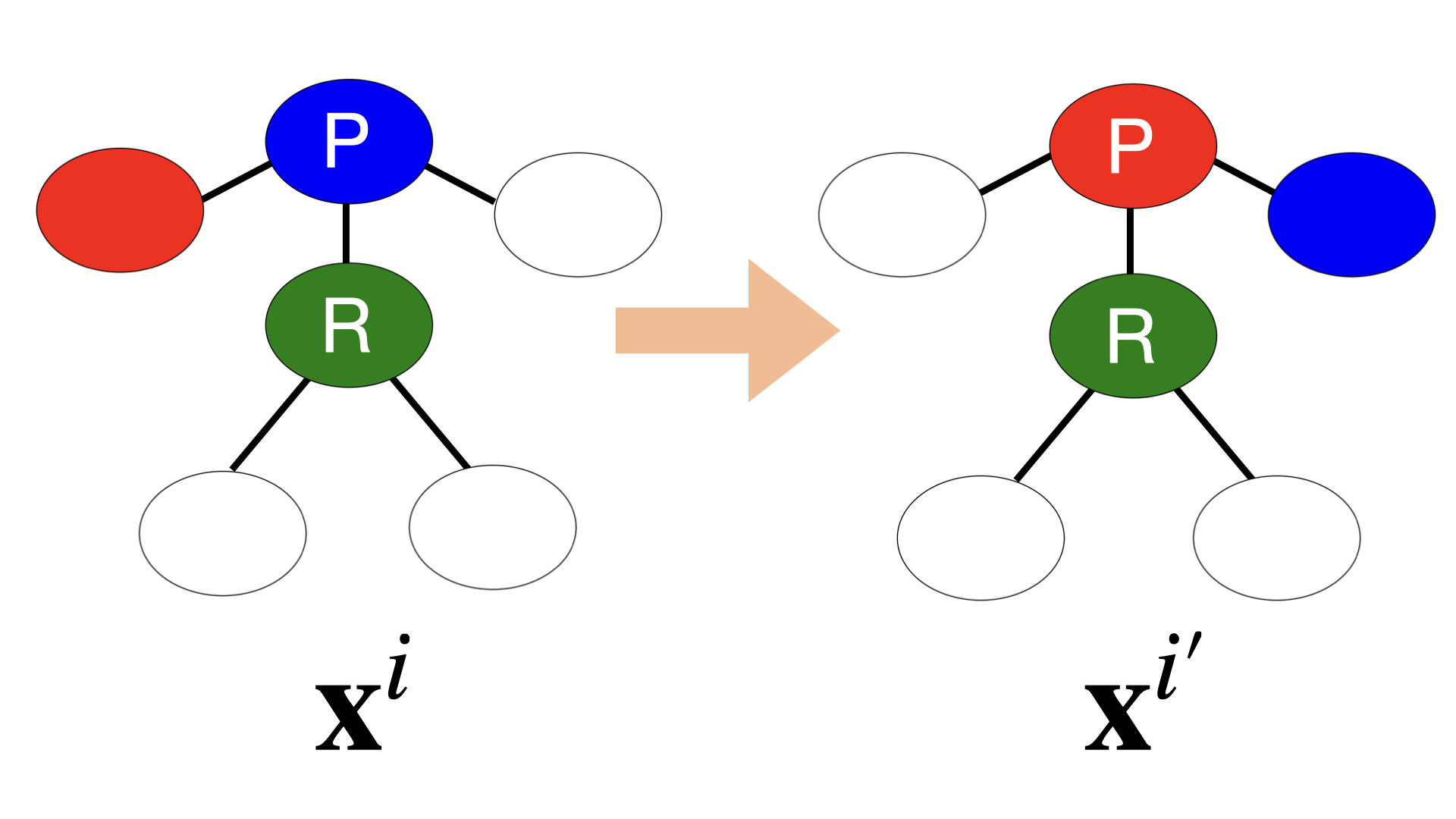}
\end{subfigure}
\begin{subfigure}{0.35\linewidth}
    \includegraphics[width=\linewidth]{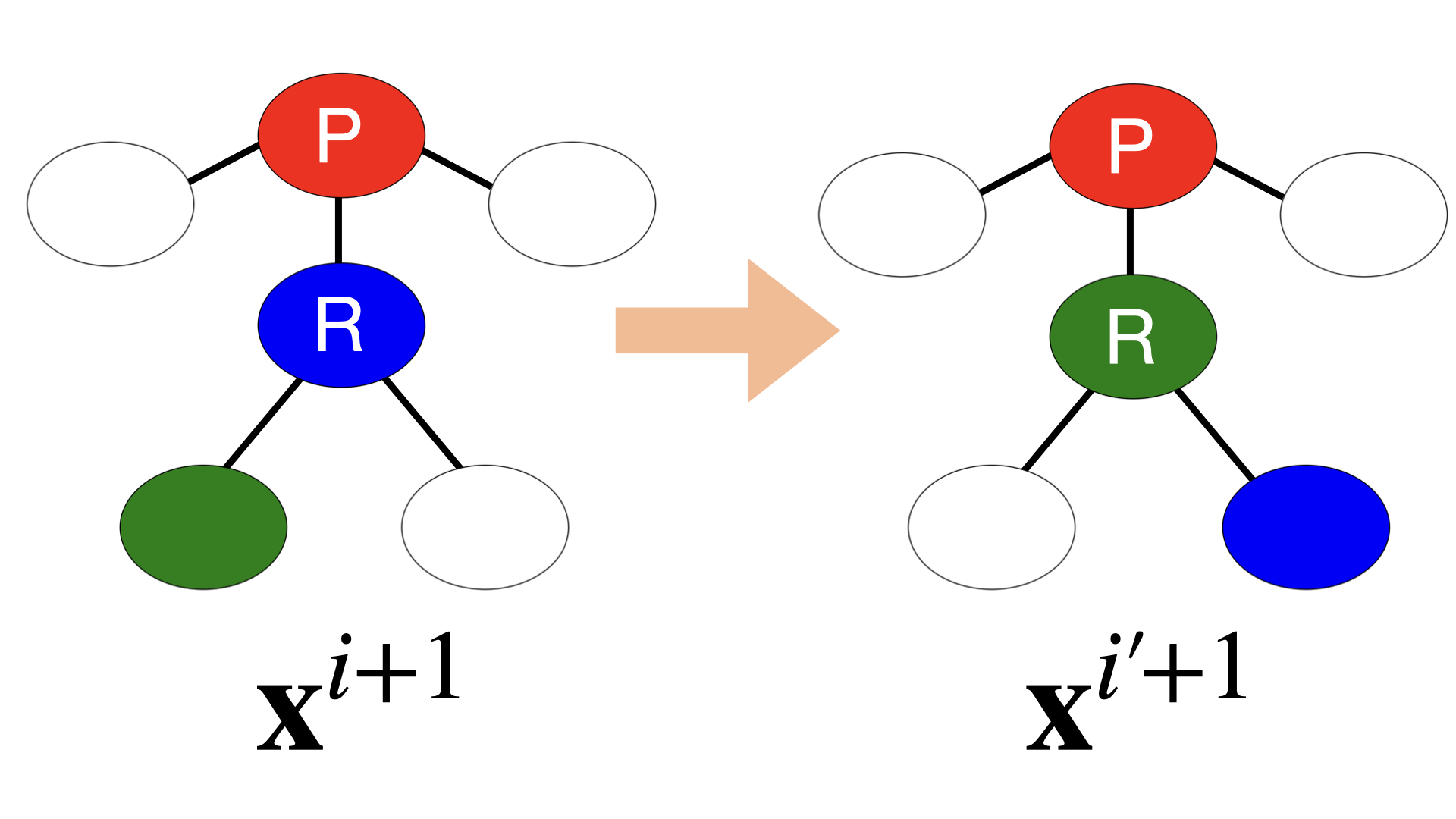}
\end{subfigure}

\begin{subfigure}{0.8\linewidth}
    \includegraphics[width=\linewidth]{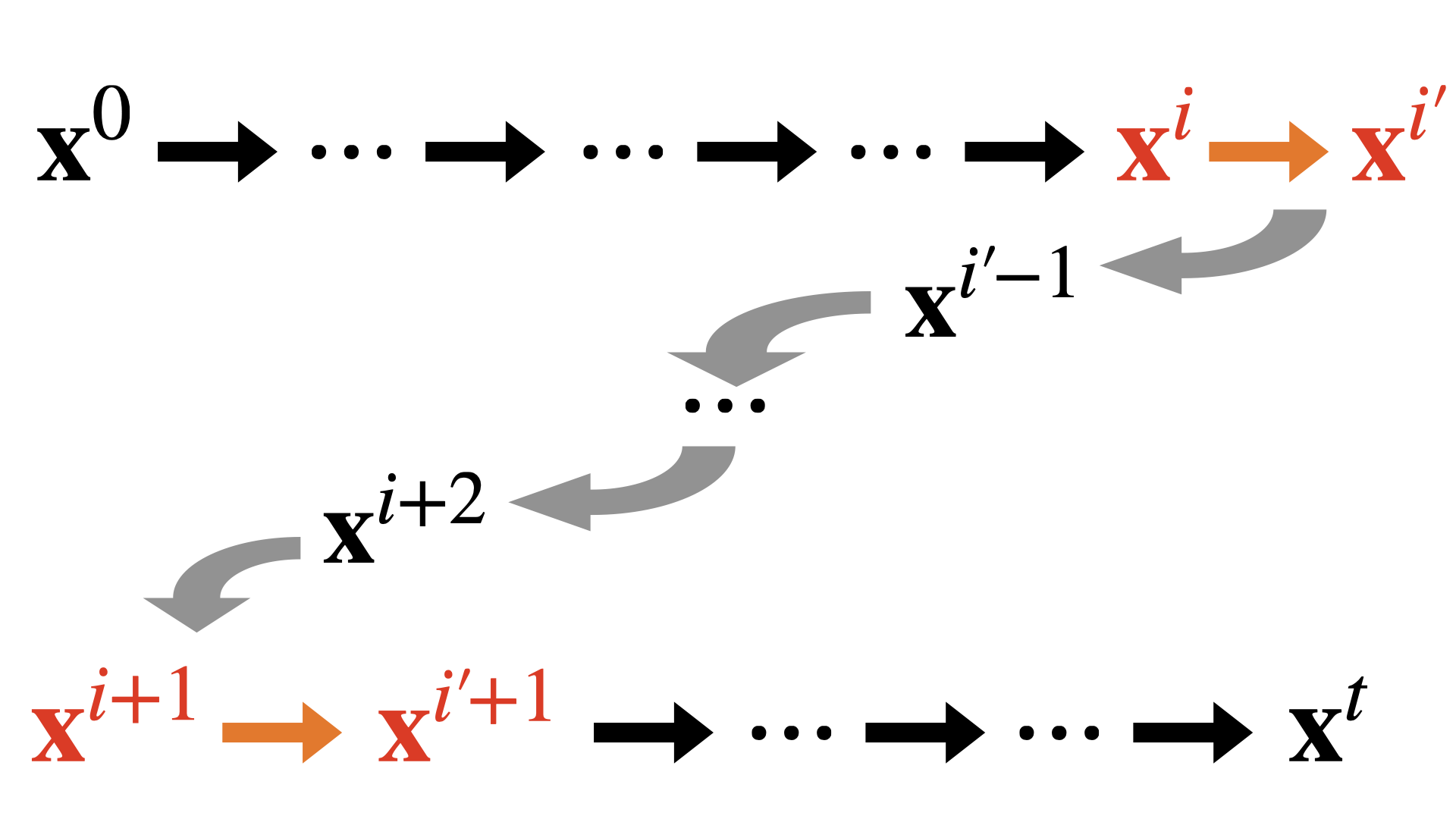}
\end{subfigure}

\caption{Z-transform for a repeated LL transition shape. Removed transitions are colored in pink and added transitions are colored in orange.} \label{fig:z-lemma-ll}
\end{figure}

\begin{proof}
Let $\calX$ be an optimal traversal, and assume in contradiction that a transition shape repeats. We construct a new traversal $\calY$ with $\time{\calY}\le \time{\calX}$, that reduces the number of transition shape repetitions by one. We will show how it works for LL transitions, the other 11 cases are provided in Appendix~\ref{sec:z-transform-three-robot-shapes}.

Assume there exist $i<i'$ and $R\in V$, such that the transitions $(\xx^i, \xx^{i+1})$ and $(\xx^{i'},\xx^{i'+1})$ are both LL-transitions that enter the sub-tree rooted at $R$. Then consider traversal $\calY$ that is defined as follows. Traversal $\calY$ follows $\calX$ from initial configuration $\xx^0$ up until $\xx^i$. It then moves to configuration $\xx^{i'}$, and then follows $\calX$ in reverse, to configuration $\xx^{i'-1}$, and keeps going in reverse until reaching configuration $\xx^{i+1}$. It then proceeds to configuration $\xx^{i'+1}$, and then follows $\calX$ from this point onward. Since $\calY$ consists of the same set of configurations in a different order, it visits all vertices, and is therefore a valid traversal.

Note that $(\xx^i,\xx^{i'})$ and $(\xx^{i+1},\xx^{i'+1})$ are valid transitions specifically for the L-shaped configurations, as depicted in Figure~\ref{fig:z-lemma-ll}. This is not the case in all transition shapes. Indeed, for an \flip{P}V transition, $(\xx^{i+1},\xx^{i'+1})$ is not a valid transition. Nevertheless, we the robots can get from $\xx^{i+1}$ to $\xx^{i'+1}$ by going through a configuration where all robots are at $R$. While this increases the traversal time by $1$, we note that in this case also $\xx^{i}=\xx^{i'}$, and therefore the overall traversal time is maintained. This happens in all 12 cases.
\end{proof}


\section{Applications}
\label{sec:apps}
In this section we discuss some applications of $\MRFGC$. We may consider the following applications:
\begin{enumerate}
    \item \label{rule:communication} \emph{Local communication.}  All robots must be within some small communication distance.
    \item \label{rule:collision} \emph{Collision avoidance.} Robots cannot occupy the same vertex.
    \item \label{rule:weight-width} \emph{Passage width/material load capacity.} Edges of the graph are parameterized with material load capacity and a passage width. These could restrict the set of robots that can go through an edge simultaneously, that is, reduce the set of valid transitions.
    \item \label{rule:guards} \emph{Guards.} For robots of a certain type to cross and edge, the end-points must be occupied by robots of some type. This can model securing a passage before a more valuable robot can cross it.
    \item \label{rule:coverage} \emph{Cleaning.} The graph must be covered by robots of a certain type. For example in cleanup tasks, one type of robots may be used for cleaning, and another for grinding the garbage.
\end{enumerate}



Note that application~\ref{rule:collision} is not collapsible, and hence the $\PTAS$ algorithm does not apply in this setting. Also, the $\PTAS$ algorithm for $3$ robots assumes homogeneous robots, and hence only the first application applies.

\paragraph{Preliminary Experimental Results.} We tested the performance of our approach (with some heuristics simplifications) on floor plans of several large hotel buildings, and obtained consistent improvement in traversal time compared to the current state-of-the-art approach by~\cite{sinay2017maintaining}.


\appendix

\bibliography{references}

\section{Examples of Formations and Transpositions}
\label{sec:examples}
In this section, we provide an illustrative example of formations and transpositions. We consider two robot types: \emph{routers} (colored in red) and  \emph{cleaners} (colored in blue). Formally, $M=\{r,b\}$. For ease of exposition, we assume there is one router and two cleaners, namely, $k_r=1,k_b=2$. We require the cleaners to stay connected to the router. In more general, we may consider a network of routers that always forms a connected subgraph, and cleaners should always stay within some range from a router. 

\begin{figure}[htb]
\centering
\begin{subfigure}{0.3\linewidth}
    \includegraphics[width=\linewidth]{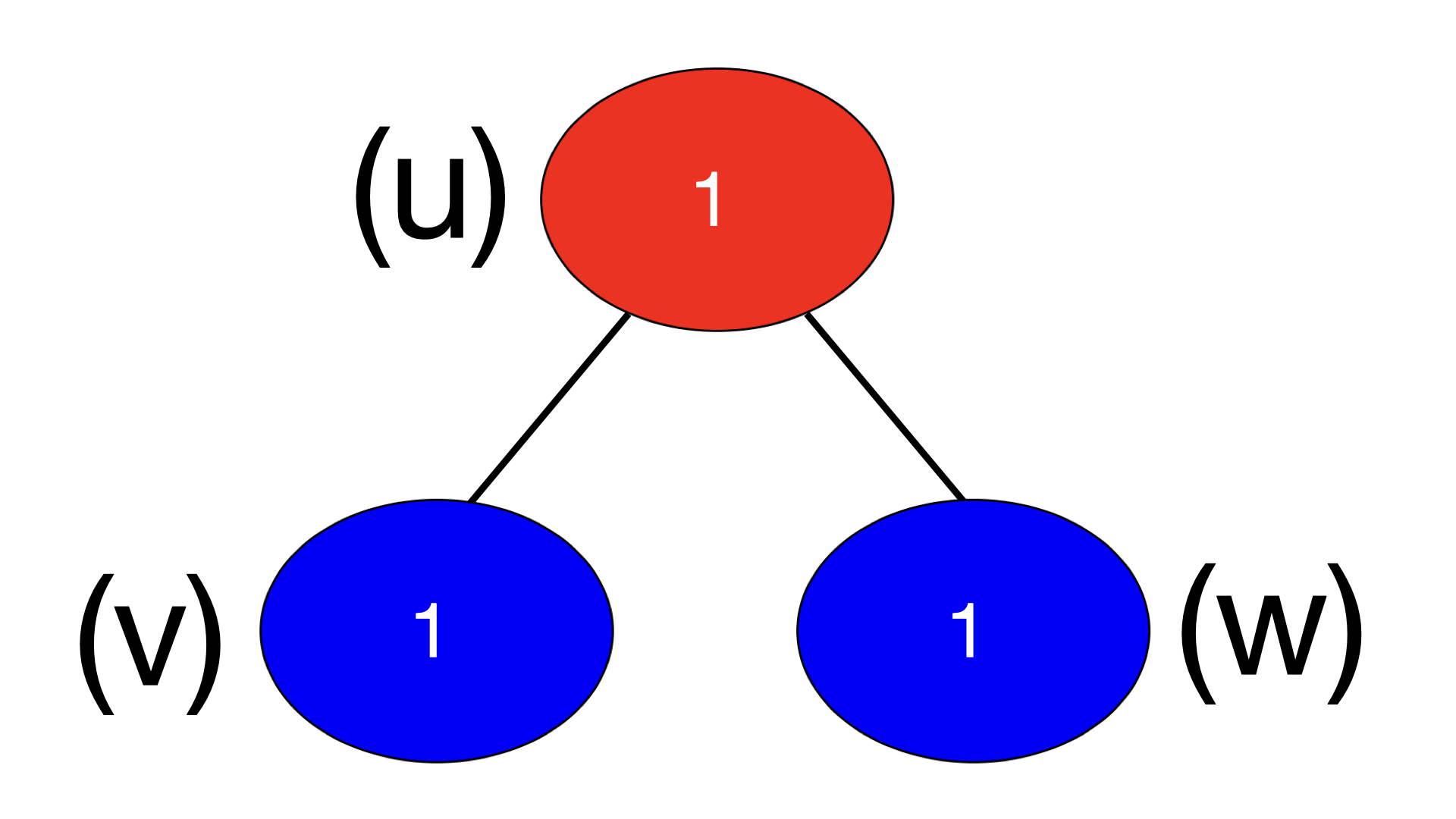}
    \caption*{($\alpha$)}
\end{subfigure}
\begin{subfigure}{0.3\linewidth}
    \includegraphics[width=\linewidth]{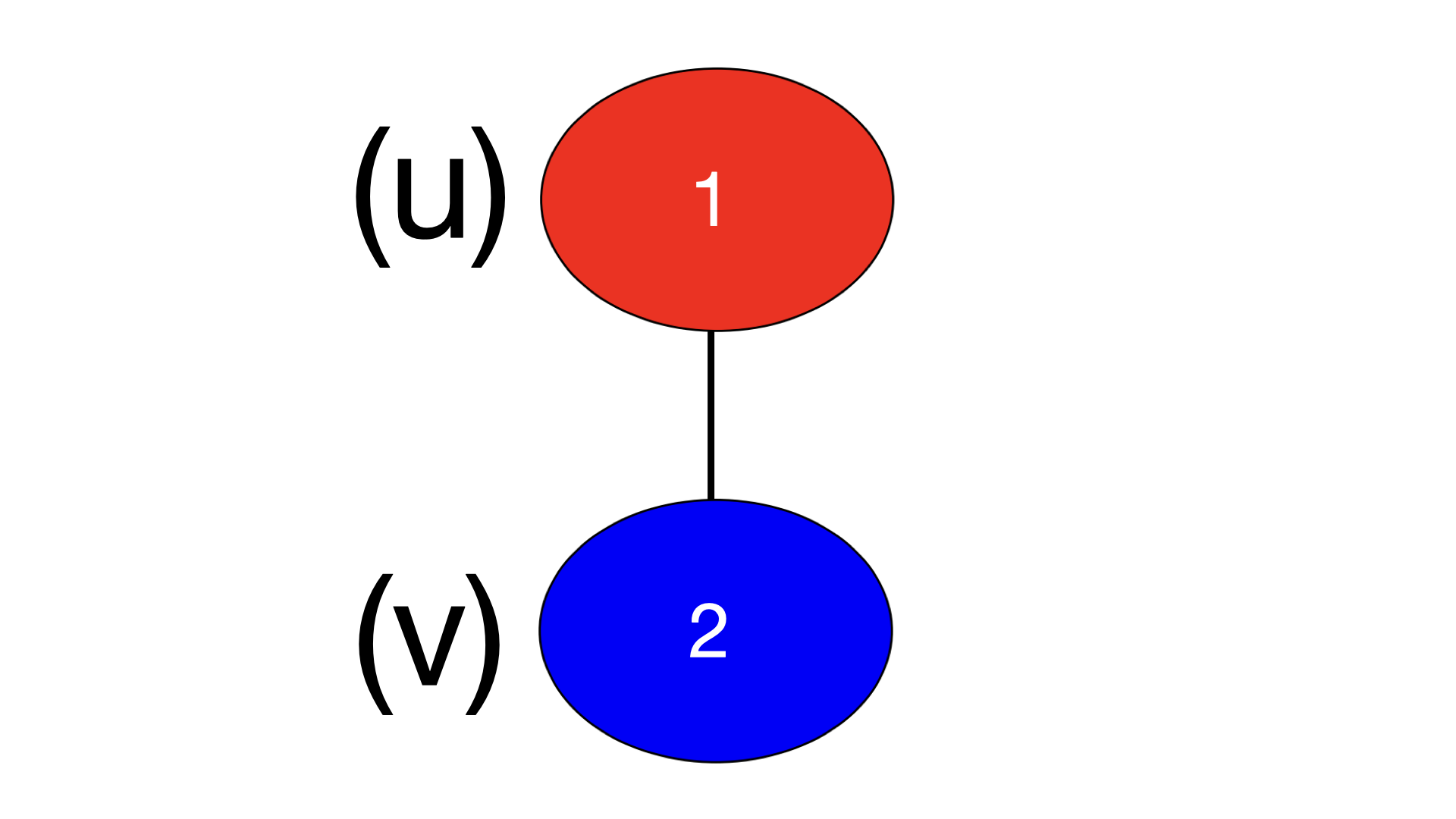}
    \caption*{($\beta$)}
\end{subfigure}
\begin{subfigure}{0.3\linewidth}
    \includegraphics[width=\linewidth]{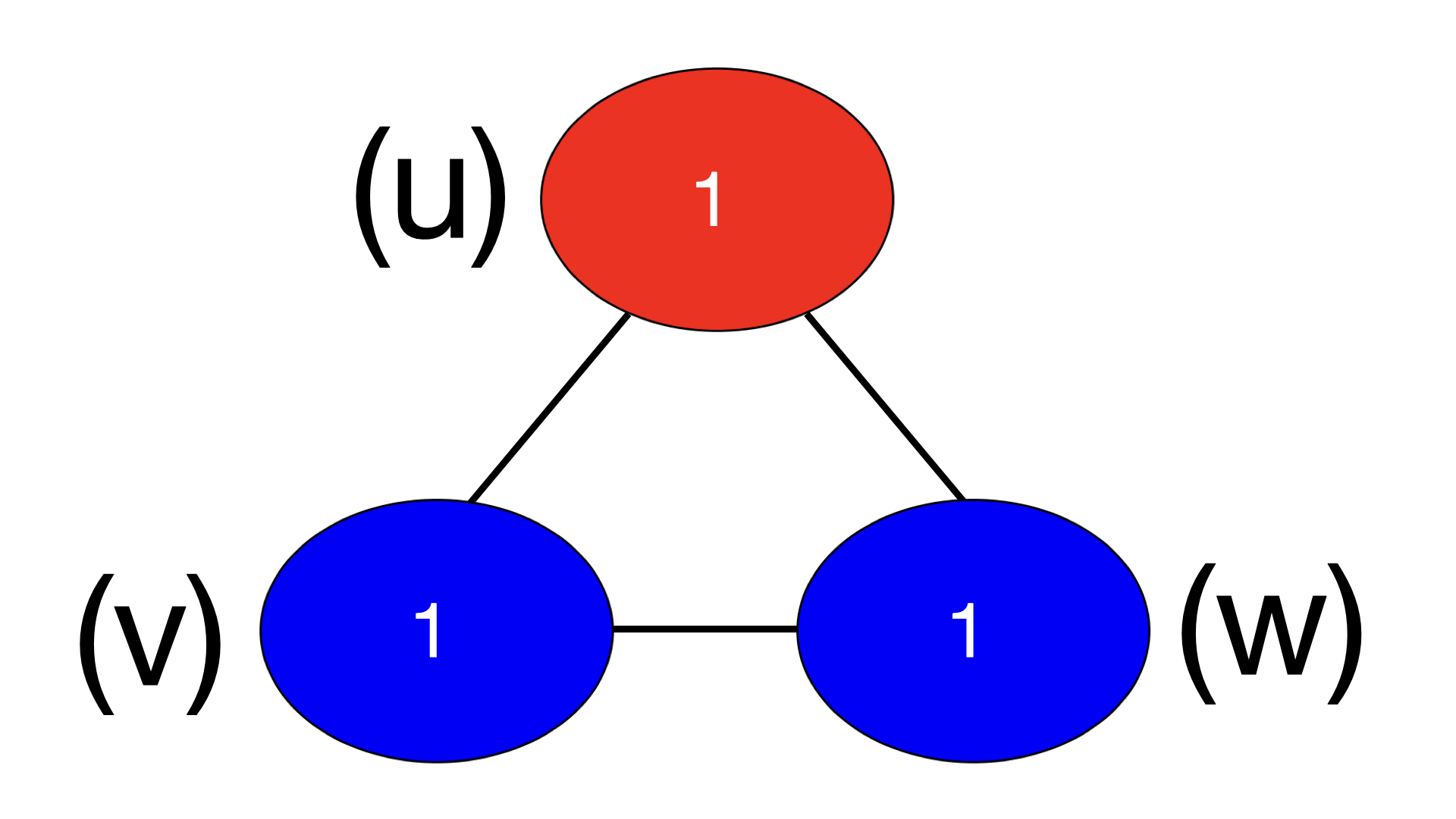}
    \caption*{($\gamma$)}
\end{subfigure}

\caption{Router-Cleaner example formations.} \label{fig:example-formations}
\end{figure}

Figure~\ref{fig:example-formations} depicts three possible router-cleaner formations with $k_r=1,k_b=2$. For instance, consider the $\gamma$ formation $\langle G_\gamma, \xx_\gamma \rangle$. The graph $G_\gamma=(V_\gamma,E_\gamma)$ is given as $V_\gamma=\{u,v,w\}$, and $E_\gamma=\{\{u,v\}, \{v,w\}, \{u,w\}\}$. The function $\xx_\gamma: V_\gamma \times M \rightarrow \NN$ then specifies how many routers and cleaners occupy each vertex. Namely, $\xx_\gamma(u,r)=\xx_\gamma(v,b)=\xx_\gamma(w,b)=1$, and $\xx_\gamma(\cdot,\cdot)=0$ otherwise.

\begin{figure}[htb]
\centering
\includegraphics[width=\linewidth]{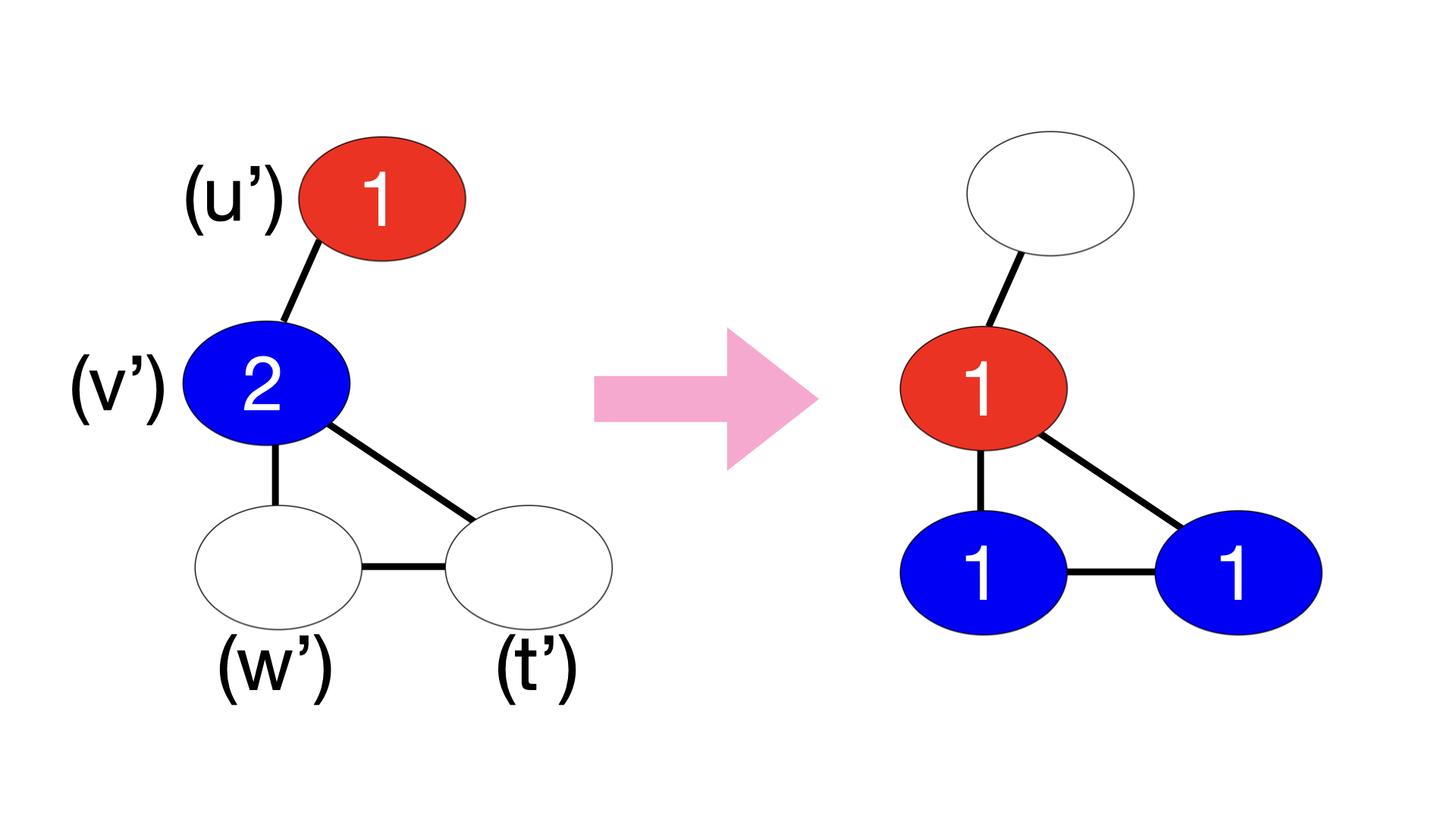}
\caption{($\beta \rightarrow \gamma$) Transposition}
\label{fig:example-transposition}
\end{figure}

Figure~\ref{fig:example-transposition} depicts a transposition from a $\beta$ formation to a $\gamma$ formation, $\langle G_{\{\beta, \gamma\}}, \{\xx_\beta, \xx_\gamma\} \rangle$. The graph $G_{\{\beta, \gamma\}}=\langle V_{\{\beta, \gamma\}}, E_{\{\beta, \gamma\}}\rangle$ consists of $V_{\{\beta, \gamma\}}=\{u',v',w',t'\}$ and $E_{\{\beta, \gamma\}}=\{\{u',v'\}, \{v',w'\}, \{w',t'\}, \{v',t'\}\}$. The configuration $\xx_\beta$ admits $\xx_\beta(u',r)=1, \xx_\beta(v',b)=2$ and zero otherwise. Indeed, $\xx_\beta$ is in $\beta$-form. Formally, we may consider the graph monomorphism $\phi: G_\beta \rightarrow G_{\{\beta,\gamma\}}$, that maps $u \mapsto u', v\mapsto v'$. Similarly, configuration $\xx_\gamma$ is in $\gamma$ form, considering the monomorphism $u\mapsto v', v\mapsto w', w\mapsto t'$. Lastly, the $\gamma$ configuration can be reached from the $\beta$ configuration by moving each robot along an edge. The router moves from $u'$ to $v'$, and the cleaners move from $v'$ to $w'$ and $t'$.

\begin{figure}[htb]
\centering
\includegraphics[width=\linewidth]{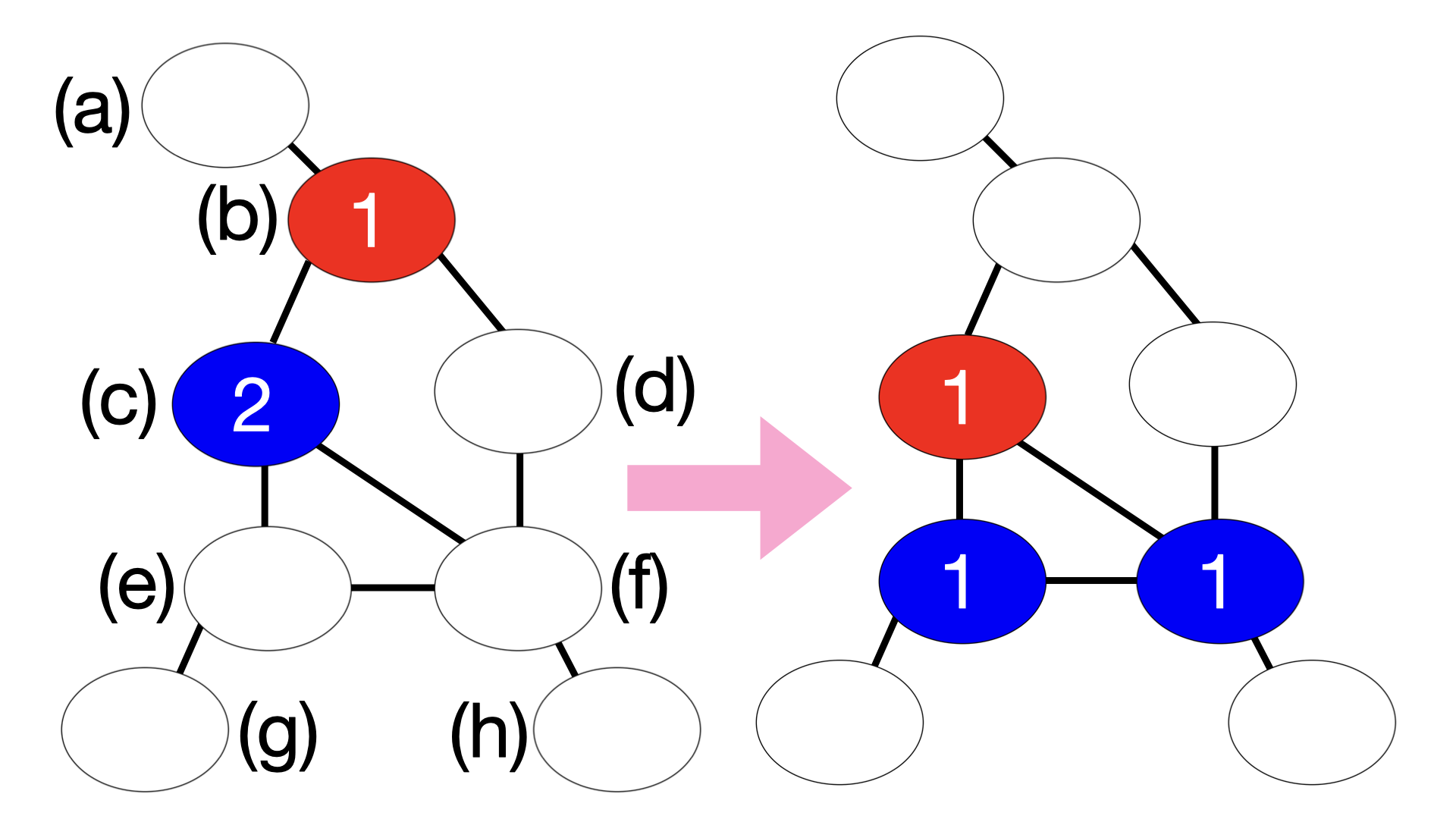}
\caption{($\beta \rightarrow \gamma$) Transition}
\label{fig:example-transition}
\end{figure}

Finally, Figure~\ref{fig:example-transposition} depicts a $\beta \rightarrow \gamma$ transition in a given graph $G=\langle V, E \rangle$, where $V=\{a,b,c,d,e,f,g,h\}$ and $E=\{ \{a,b\}, \{b,c\}, \{b,d\}, \{c,e\}, \{c,f\}, \{d,f\}, \{e,g\}, \{e,f\}, \{f,h\}\}$. Indeed, consider the monomorphism that maps $u',v',w',t'$ to $b,c,e,f$ respectively.

\section{Nice Tree Decomposition}
\label{sec:nice-tree-decomp}
In this section, we recall the definition of a nice tree decomposition~\cite{kloks1994treewidth}. We follow the formulation in~\cite{van2015dynamic}. We note that our algorithm can work directly on any tree decomposition. Nevertheless, the description is more intuitive when working with nice tree decomposition.

A tree decomposition is converted into a nice tree decomposition, in order to limit the structure into a small set of possible transitions between bags (not to be confused with robot transitions). The three different transitions are:
\begin{itemize}
    \item An \textsf{introduce} bag introduces a new vertex. It has a single child that has the same bag with the new vertex excluded.
    \item A \textsf{forget} bag removes a vertex. It has a single child that has the same bag with the forgotten vertex included.
    \item A \textsf{join} bag joins two bags with equal contents. It has two children with the exact same bag.
\end{itemize}
Notably, some works also consider an \textsf{introduce edge} bag, that contains the same vertices as its child, but is labeled with the edge it introduces. We did not find this additional refinement beneficial in our approach.

In Section~\ref{sec:fpt-algorithm}, we observe that signatures of a given traversal at a bag and its child in a nice tree decomposition are related. The definitions of reduce, lift and combine, specify these relations for introduce, forget and join nodes respectively. This simplifies both the description and the analysis of the algorithm.

\section{Omitted Proofs for $\MRFGC$}
\label{sec:omitted-proofs}
In order to prove the correctness of our algorithm, we first define \emph{partial traversals}. These are similar to signatures, but instead of focusing on the traversal pattern at a given bag $j$, a partial traversal keeps track of all configurations that intersect $V_\downarrow (j)$.

\begin{Definition}
Let $\calX=(\xx^0,\ldots,\xx^t)$ be a traversal and $j\in J$. The \emph{partial traversal} of $\calX$ at $j$, denoted $\parti{\calX}{j}$, is the sequence $\calY=(\yy^0,\ldots,\yy^t) \in \configs_{B_j} \cup \{\uparrow\}$, where for each $0\le i \le t$:

\begin{align*}
\yy^i = \begin{cases}
\uparrow & \activated{\xx^i} \subseteq V_\uparrow (j) \\
\xx^i    & \text{otherwise} \\
\end{cases}
\end{align*}

The \emph{condensed form} of $\parti{\calX}{j}$, denoted $\cparti{\calX}{j}$, is obtained from $\parti{\calX}{j}$ by replacing any consecutive repetition of the same element with a single occurrence of that element.

The \emph{cost} of $\cparti{\calX}{j}$, denoted $\cost(\cparti{\calX}{j})$, is the number of configurations whose active vertices intersect $V_\downarrow (j)$.
\end{Definition}

Evidently, $\lift{\cparti{\calX}{j}}{B_j}=\cproject{\calX}{j}$, so we can get the projection on $j$ from the partial solution at $j$. We will show that it is also possible to get the opposite: we can reconstruct a partial solution at $j$ from a condensed sequence $\pattern^j_\ell$ on $j$ that appears in an updated $\table_j$ and has finite cost. Ultimately, for $\treeroot \in J$, we can reconstruct a traversal from any condensed sequence with finite cost.

Indeed, this is achieved by calling \textsf{Reconstruct}~\ref{algorithm:reconstruct}. This is a recursive algorithm that reconstructs a partial solution at bag $j\in J$ for the $\ell$\textsuperscript{th} signature, from a reconstructed partial solution of $j$'s children, by recursively calling \textsf{Reconstruct} for the child tables, at the entries specified by the pointers.

\begin{algorithm}[ht]
\caption{\textsf{Reconstruct}}\label{algorithm:reconstruct}
\begin{algorithmic}
    \REQUIRE Bag index $j\in J$, row index $\ell$, where $\cost^j_\ell < \infty$;
    \ENSURE A (candidate) condensed partial solution $\calY^j_\ell$;
    
    \hrulefill

    \IF{$j$ is a \codebox{leaf} node}
        \RETURN $(\uparrow)$;
    \ENDIF

    \IF{$j$ is an \codebox{add} node, with $B_j=B_{j'}\cup\{v\}$}
        \STATE Set $\ell'=\pointers^j_\ell$;
        \RETURN $\substitute{\pattern^j_\ell}{\textsf{Reconstruct}(j',\ell')}$;
    \ELSIF{$j$ is a \codebox{forget} node, with $B_j=B_{j'}\setminus\{v\}$}
        \STATE Set $\ell'=\pointers^j_\ell$;
        \RETURN $\textsf{Reconstruct}(j',\ell')$;
    \ELSIF{$j$ is a \codebox{join} node, with children $j',j''$ in $\calT$}
        \STATE Set $(\ell',\ell'')=\pointers^j_\ell$;
        \STATE Set $\calY^{j'}_{\ell'}=\textsf{Reconstruct}(j',\ell')$;
        \STATE Set $\calY^{j''}_{\ell''}=\textsf{Reconstruct}(j'',\ell'')$;
        \RETURN $\substitute{\pattern^j_\ell}{\calY^{j'}_{\ell'},\calY^{j'}_{\ell'}}$;
    \ENDIF
\end{algorithmic}
\end{algorithm}

\begin{Lemma}
\label{lemma:reconstruction}
After calling $\textsf{UpdateAllTables}$, let $j\in J$ and $1\le \ell \le |\poss{j}|$ such that $\cost^j_\ell < \infty$. Let $\calY^j_\ell = \textsf{Reconstruct}(j,\ell)$. Then one of two must hold:
\begin{itemize}
    \item $\calY_\ell^j$ is a condensed partial solution at $j$ with $\cost(\calY_\ell^j)=\cost_\ell^j$.
    \item Bag $j\neq \treeroot$ is \emph{not} the root of $\calT$, and there exist no traversal $\calX$ with $\cproject{\calX}{j}=\pattern_\ell^j$.
\end{itemize}
\end{Lemma}

\begin{proof}
We prove by induction. If $j$ is a leaf, $\calY^j_\ell=(\uparrow)=\cparti{\calX_0}{j}$ for any traversal $\calX_0$, and its cost is $\cost(\calY^j)=\cost^j_\ell=0$. Fix some $\ell$, let $j\in J$ and assume the above holds for each $j'$ child of $j$ and each $\ell'$.


By induction hypothesis, there exist traversals $\calX^{j'}_{\ell'}$ for each child $j'$ of $j$, such that $\calY^{j'}_{\ell'}:=\textsf{Reconstruct}(j',\ell')=\cparti{\calX^{j'}_{\ell'}}{j'}$
with $\cost(\cparti{\calX^{j'}_{\ell'}}{j'})=\cost^{j'}_{\ell'}$. If $j=\treeroot$ is the root in $\calT$, we may define the following traversal $\calZ^j_\ell$:
$\cproject{\calZ^j_\ell}{j}=\pattern^j_\ell$, and $\cparti{\calZ^j_\ell}{j'}=\calX^{j'}_{\ell'}$ for each child $j'$ of $j$ where $\ell'$ is given by the pointer. Since $j=\treeroot$, $\pattern^j_\ell$ contains no $\uparrow$'s, and therefore $\calZ^j_\ell$ is a traversal with $\cproject{\calZ^j_\ell}{j}=\pattern^j_\ell$.

Therefore, we are left with the case of $j\neq \treeroot$ and there exist a partial solution with pattern $\pattern_\ell^j$, denoted $\calX_\ell^j$. In this case, we construct the following traversal $\calZ^j_\ell$: $\reduce{\calZ^j_\ell}{V_\uparrow (j)}=\reduce{\calX^j_\ell}{V_\uparrow (j)}$, $\cproject{\calZ^j_\ell}{j}=\pattern^j_\ell$, and $\cparti{\calZ^j_\ell}{j'}=\calX^{j'}_{\ell'}$ for each child $j'$ of $j$ where $\ell'$ is given by the pointer. Therefore, $\cparti{\calZ^j_\ell}{j}=\calY^j_\ell$.

As for the cost, first observe that since the configurations added in $\calX_\ell^j$ do not add to the cost at $j$, as they are all contained in $V_\uparrow (j)$.

Next, we consider each node type: if $j$ is an add node, the cost is the same since $V_\downarrow (j)=V_\downarrow(j')$, where $j'$ is the child of $j$, and the cost is indeed copied; for a forget node, we add to the cost all the forgotten configurations at $j$; for a join node, note that $V_\downarrow(j)=V_\downarrow (j')\cup V_\downarrow (j'')$, and no forgotten configuration can be double-counted. Indeed, by the definition of tree decomposition, any path from $V_\downarrow(j')$ to $V_\downarrow(j'')$ must intersect bag $j$. If a configuration appears in the pattern of $j'$ and of $j''$, there exist formations $\alpha',\alpha''$ and graph monomorphisms $\phi',\phi''$ mapping $G_{\alpha'},G_{\alpha''}$ to $G$ respectively. Therefore, $\phi'(G_{\alpha'})=\phi''(G_{\alpha''})$ is a connected sub-graphs of $G$ that intersects both $j'$ and $j''$, and therefore, it must intersect bag $j$. Therefore, this configuration is not forgotten neither at $j'$ nor at $j''$. Therefore, by induction, the cost of $\calZ^j_\ell$ is precisely the number of forgotten configurations, that is, configurations whose active vertices are in $V_\downarrow (j)$.
\end{proof}

Lemma~\ref{lemma:reconstruction} ensures that any row with $\cost<\infty$ is either \emph{feasible}, that is, applying $\textsf{Reconstruct}$ will restore a partial solution with this cost, or \emph{infeasible}, in which case there is no traversal of $G$ with this pattern. Since the latter only happens for tables that are not for the root of $\calT$, we are ensured to be able to reconstruct a traversal with the specified cost from any entry of the root table with finite cost.

Therefore, it remains to prove that there will exist a row in the root table with a cost that equals the optimal traversal time. Essentially, this follows from the observation that if two traversals $\calX,\calY$ have the same signature at $j$, we can replace $\reduce{\calX}{V_\downarrow (j)}$ with $\reduce{\calY}{V_\downarrow (j)}$, and get a valid traversal. Therefore, always picking the child signatures with minimal cost must lead to an optimal traversal, and tie-breaking can be done arbitrarily.

\begin{Lemma}
\label{lemma:optimal-survives}
Let $j\in J$ and let $\calX$ be an optimal traversal. Then there exist a row $\ell$ such that $\pattern^j_\ell=\cproject{\calX}{j}$, and $\cost^j_\ell=\cost(\cparti{\calX}{j})$.
\end{Lemma}

\begin{proof}
We prove by induction on the tree $\calT$. For the leaves, this is trivially true as the projection is $(\uparrow)$ with a cost of 0. In any node type, the signature $\cproject{\calX}{B_j}$ is considered in $\textsf{enumerate\_patterns}$ by Lemma~\ref{lemma:identify-patterns}, say in row $\ell$.

If $j$ is an add node with child $j'$, then $\cproject{\calX}{j'}=\reduce{\pattern^j_\ell}{B_{j'}}$, and by induction, it appears in the table, in row $\pointers^j_\ell=\ell'$. The cost at $j$ is correct by induction, as it should be the same cost as $j'$, since no vertex is forgotten.

If $j$ is a forget node with child $j'$, then $\lift{\cproject{\calX}{j'}}{B_j} = \cproject{\calX}{j}$. By induction, a row $\ell'$ with $\pattern^{j'}_{\ell'}=\cproject{\calX}{j'}$ appears in $\table_{j'}$, and $\cost^{j'}_{\ell'}=\cost(\cparti{\calX}{j})$. Row $\ell'$ also minimizes the objective function. Indeed, assume there exist $\ell''\in L'$ with a smaller $\cost^j_\ell$. Then we can construct traversal $\calY$ such that $\cproject{\calY}{j}=\pattern^j_\ell$, $\cproject{\calY}{j'}=\pattern^{j'}_{\ell'}$, $\reduce{\calY}{V_\uparrow (j)} = \reduce{\calX}{V_\uparrow (j)}$, and $\lift{\calY}{V_\downarrow (j')}=\lift{\textsf{Reconstruct}(j',\ell')}{V_\downarrow (j')}$. It is a valid traversal in time smaller then $\time{X}$ in contradiction. While $\textsf{UpdateAllTables}$ may pick a different row from $\ell'$, it will be updated with the same cost, and same projection, and will correspond to some optimal traversal as the one we defined for $\calY$.

Similarly, if $j$ is a join node with children $j',j''$, we have that $\cproject{\calX}{j}$ combines $\cproject{\calX}{j'}$ and $\cproject{\calX}{j''}$. By induction, rows $\ell',\ell''$ in tables $\table_{j'},\table_{j''}$ with such projections will appear with the right cost. Therefore $(\ell',\ell'')$ will be in $L^2$, and will minimize the objective (otherwise $\calX$ is again not optimal). While $\textsf{UpdateAllTables}$ may pick a different row from $\ell'$, it will be updated with the same cost, and same projection.
\end{proof}

As a result, we get an $\FPT$ algorithm for $\MRCGC$~\ref{algorithm:mrcgc}. First, \textsf{UpdateAllTables}~\ref{algorithm:update-all-tables} is called to update all the tables. Then, $\MRCTC$~\ref{algorithm:mrcgc} takes the row with minimal cost from $\table_\treeroot$, and reconstructs a traversal from it using $\textsf{Reconstruct}$~\ref{algorithm:reconstruct}.

\begin{algorithm}[ht]
\caption{$\MRCGC$}\label{algorithm:mrcgc}
\begin{algorithmic}
    \REQUIRE A graph $G$ with tree decomposition $(\calB,\calT)$ and maximal degree $\maxdegree\in\NN$, number of robots $k\in \NN$;
    \ENSURE An optimal traversal $\calX$;
    
    \hrulefill

    \STATE $\table_\treeroot \gets \textsf{UpdateAllTables}()$;

    \STATE $\ell \gets \textsf{get\_min\_cost}(\table_\treeroot)$;

    \RETURN $\textsf{Reconstruct}(\treeroot, \ell)$;
\end{algorithmic}
\end{algorithm}

We are now ready to prove Theorem~\ref{theorem:fpt}:
\fpt*

\begin{proof}
By Lemma~\ref{lemma:reconstruction}, any row in the root table with $\cost<\infty$ corresponds to a valid traversal with this time. Therefore, taking the row with minimal cost will yield a traversal with time $\ge$ than the optimal traversal time. Moreover, by Lemma~\ref{lemma:optimal-survives}, a row with $\cost=\timeoptimal$ exists in the root table, and therefore $\textsf{Reconstruct}$ will recover a traversal of optimal time.

As for the runtime, computing the table of each $v\in V$ takes $\calO(h(\|\formations\|,\maxdegree,\tw))$ time, and therefore, overall computing $\table_\treeroot$ takes $\calO(n\cdot h(\|\formations\|,\maxdegree,\tw))$ time. Scanning the table for a signature of minimal cost is independent of $n$, and reconstruction takes $\calO(n)$.
\end{proof}

\section{Z-transform for $k=3$ Shapes}
\label{sec:z-transform-three-robot-shapes}
\begin{figure}[htb]
\centering
\begin{subfigure}{0.35\linewidth}
    \includegraphics[width=\linewidth]{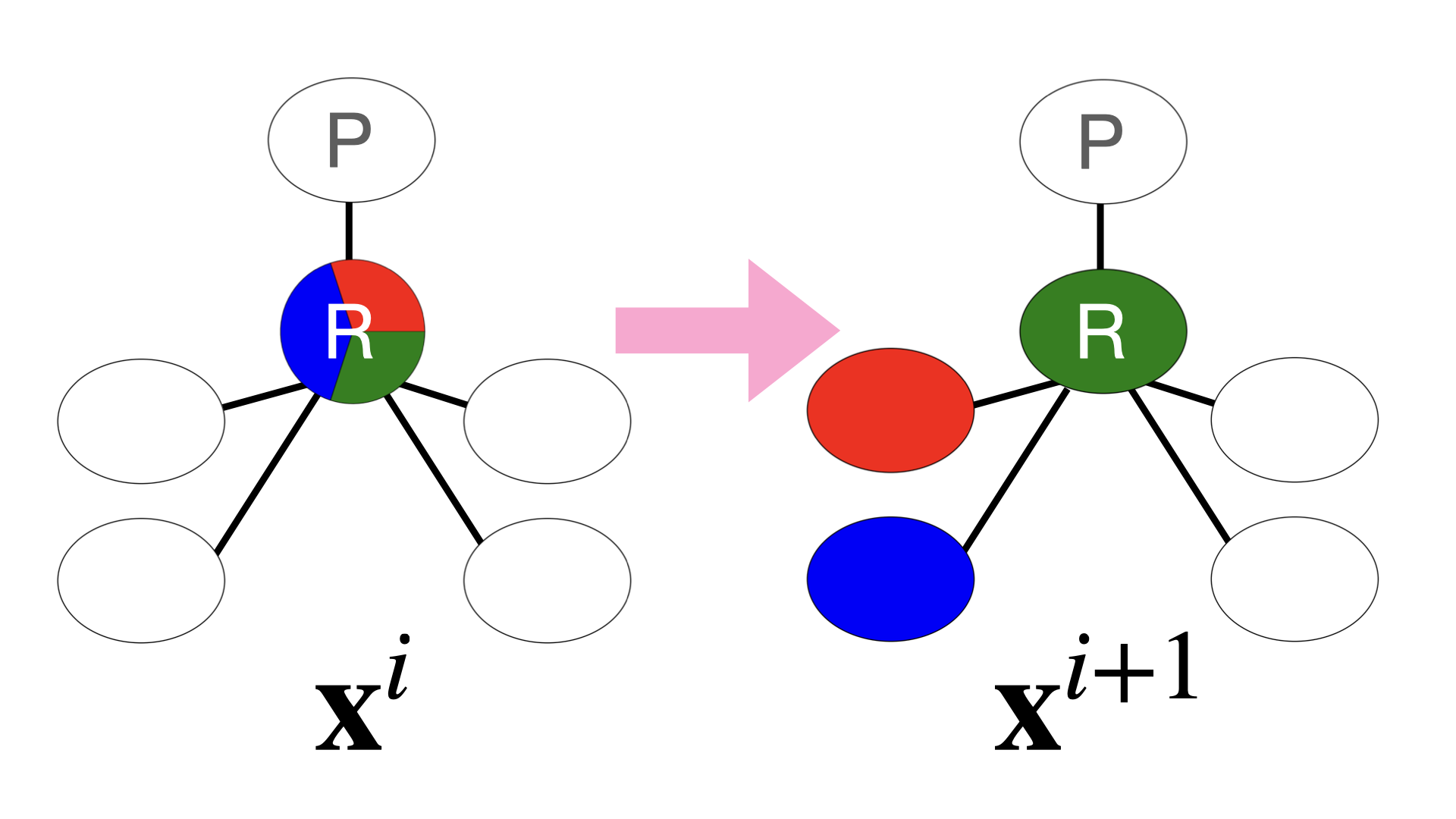}
\end{subfigure}
\begin{subfigure}{0.35\linewidth}
    \includegraphics[width=\linewidth]{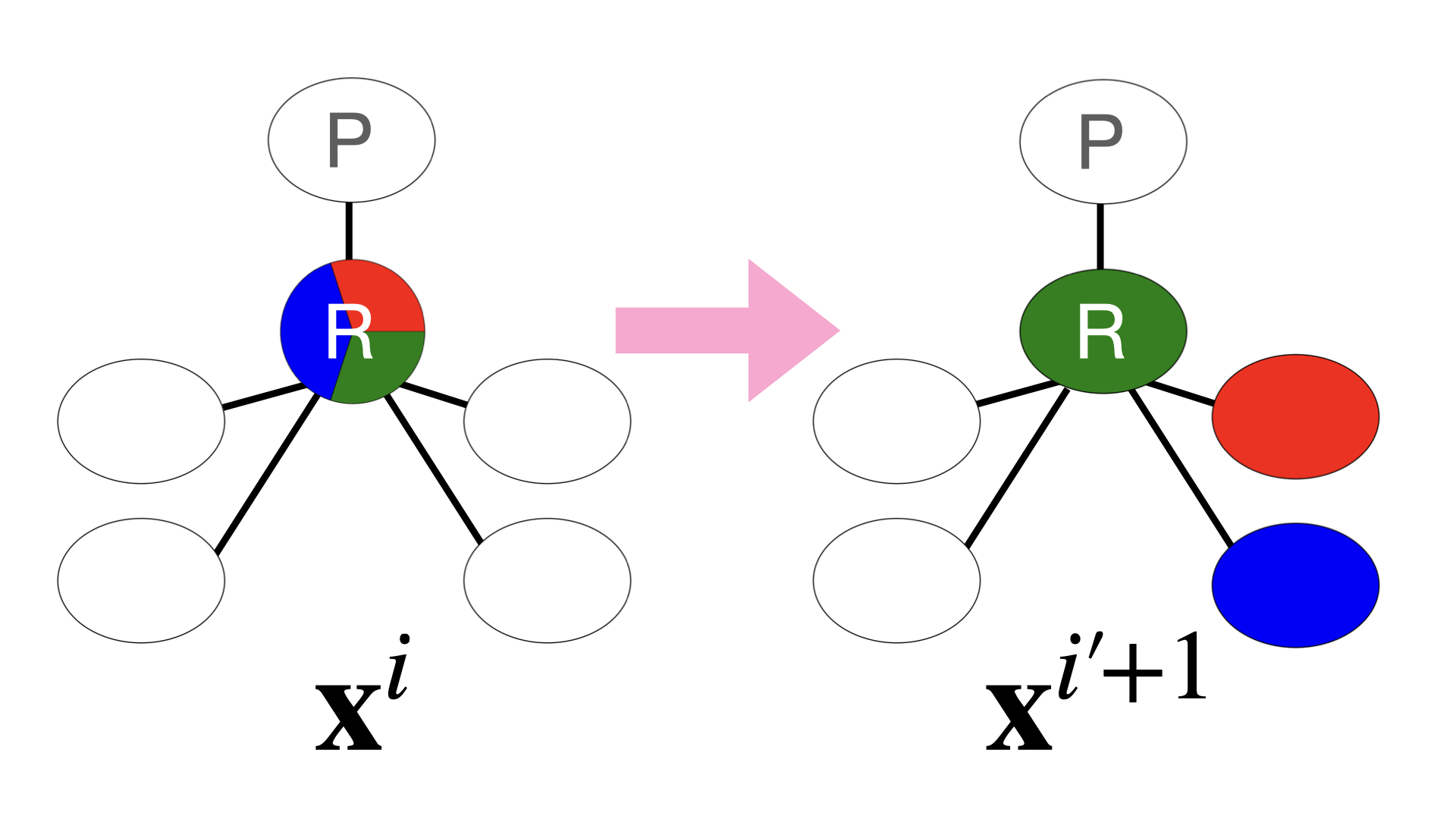}
\end{subfigure}

\begin{subfigure}{0.35\linewidth}
    \includegraphics[width=\linewidth]{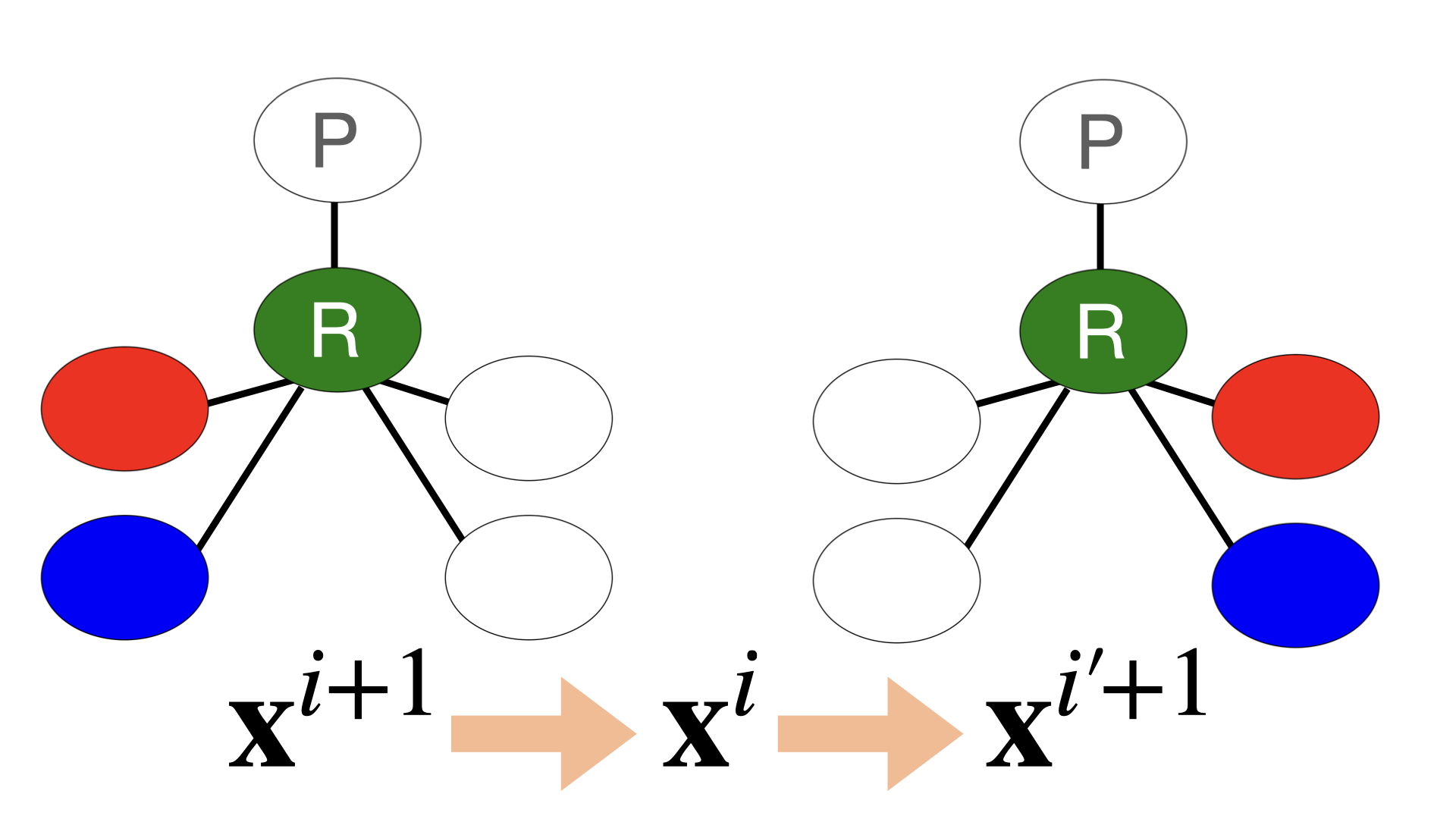}
\end{subfigure}

\begin{subfigure}{0.8\linewidth}
    \includegraphics[width=\linewidth]{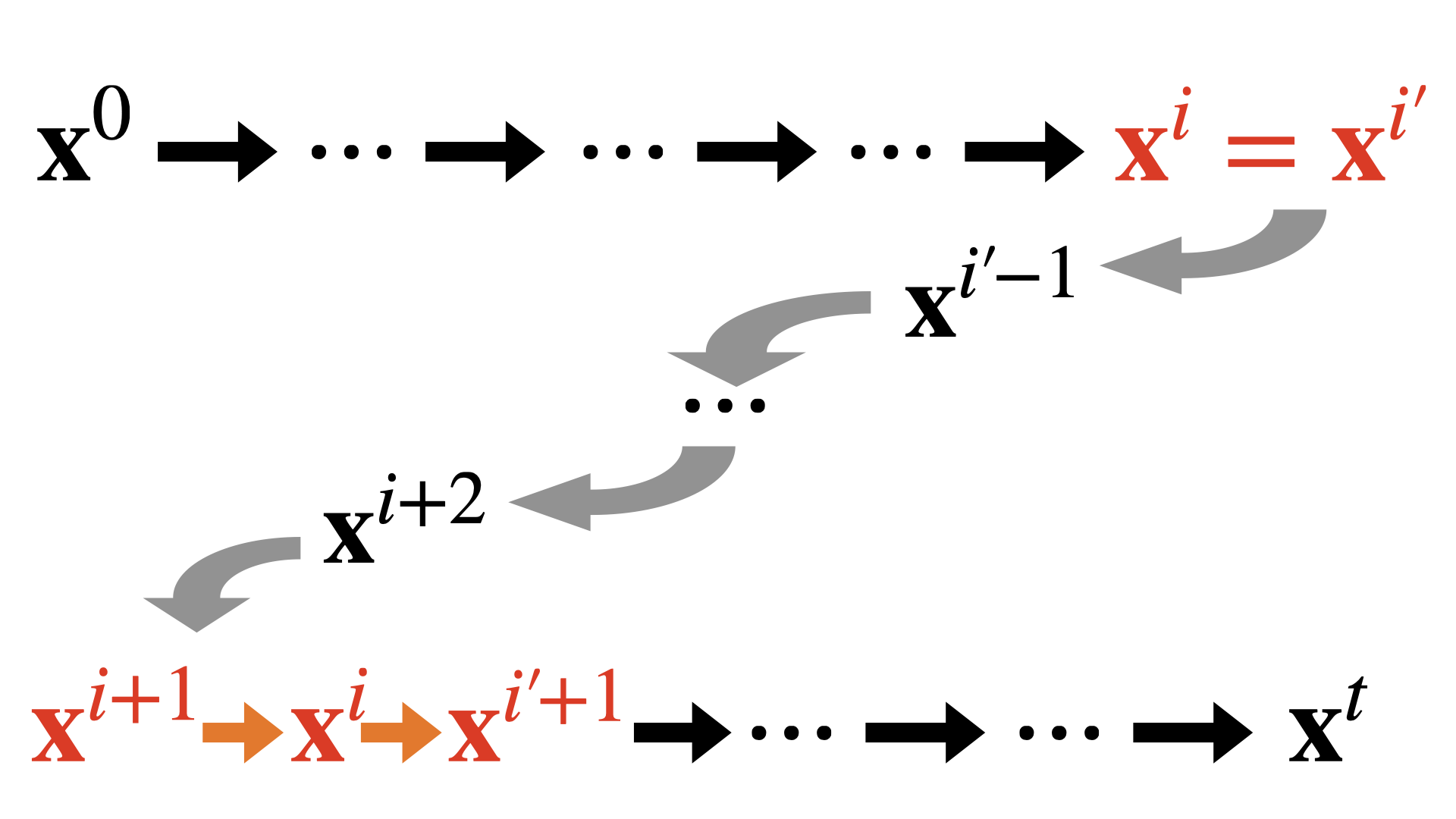}
\end{subfigure}

\caption{Z-transform for a repeated OV transition shape. Removed transitions are colored in pink and added transitions are colored in orange.} \label{fig:z-lemma-ov}
\end{figure}

In this section, we complete Proposition~\ref{prop:upper-bound-k3} by analyzing the remaining 11 types of repetitions. In Figure~\ref{fig:z-lemma-ov}, we show how to apply the Z-transform for repeated OV transition shapes. In general, any transition shape O- can use the same transformation: follow $\calX$ until reaching $\xx^i=\xx^{i'}$, then proceed in reverse to $\xx^{i+1}$, then regroup at $R$ by going to $\xx^i=\xx^{i'}$, and follow the regroup in reverse to get to $\xx^{i'+1}$. From there, follow $\calX$ until the end. The traversal consists of the same configurations in a different order.

Similarly, for all transition shapes P- and \flip{P}-, note that $\xx^i=\xx^{i'}$, and so the same transform works. First follow $\calX$ until reaching $\xx^i=\xx^{i'}$, then proceed in reverse to $\xx^{i+1}$. Lastly, get from $\xx^{i+1}$ to $\xx^{i'+1}$ by regrouping at $R$. Here, the set of configurations in the transformed traversal may strictly contain that of the original traversal, since we add an O configuration shape. Nevertheless, the number of configurations remains the same, and so traversal time is maintained.

We note that in some transition shapes the traversal time after the transformation is strictly reduced. Specifically, whenever $\xx^i=\xx^{i'}$ and we can reach from $\xx^{i+1}$ to $\xx^{i'+1}$ directly. Namely, transitions shapes \flip{P}P, \flip{P}L, PL, PP \emph{cannot} repeat in an optimal traversal.

\end{document}